\documentclass[11pt,a4paper]{article}
\usepackage{graphicx}
\usepackage{cite}
\renewcommand{\dj}{\hbox{d\hskip-1.1ex{\raise0.640ex\hbox{--}}\skip 0.70ex}}
\newcommand{\calQ}{\mathcal{Q}}
\newcommand{\calU}{\mathcal{U}}
\newcommand{\calD}{\mathcal{D}}

\newcommand{\calL}{\mathcal{L}}
\newcommand{\bra}[1]{\langle #1 |}
\newcommand{\ket}[1]{| #1 \rangle}
\newcommand{\lsim}[1]{
\setlength{\unitlength}{12pt}
\begin{picture}(1.4,1.)
\put(.7,-0.3){\makebox(0.0,1.)[t]{$<$}}
\put(.7,-0.3){\makebox(0.0,1.)[b]{$\sim$}}
\end{picture}#1}

\if@mathematic
   \def\vec#1{\ensuremath{\mathbf{#1}}}
\else      
   \def\vec#1{\ensuremath{\mathchoice{\mbox{\boldmath$\displaystyle#1$}}
                              {\mbox{\boldmath$\textstyle#1$}}
                              {\mbox{\boldmath$\scriptstyle#1$}}
                              {\mbox{\boldmath$\scriptscriptstyle#1$}}}}
\fi

\begin{document}

\begin{flushleft}
ZTF - 99/09
\end{flushleft}

\begin{center} \Huge \bf 
Nucleon strangeness as the response to a strangeness-sensitive
probe in a class of hadron models \\[1cm]
\large
D. Klabu\v{c}ar,\footnote{klabucar@phy.hr}$^{\ast}$
K. Kumeri\v{c}ki,\footnote{kkumer@phy.hr}$^{\ast}$
B. Meli\'{c}\footnote{melic@thphys.irb.hr}$^{\dagger}$ 
and  I. Picek\footnote{picek@phy.hr}$^{\ast}$
\setcounter{footnote}{0}
\end{center}
\vspace*{5mm}
\noindent
$^{\ast}$ {\small\rm Department
 of Physics, Faculty of Science, University of Zagreb, POB 162, 
 Bijeni\v{c}ka cesta 32, HR-10001 Zagreb, Croatia}\\
$^{\dagger}$ {\small\rm Theoretical Physics Division, 
Ru\dj er Bo\v{s}kovi\'{c} Institute, Bijeni\v{c}ka cesta 54, 
HR-10001 Zagreb, Croatia}

\begin{abstract}
On top of its valence quarks, the full nucleon ground state
may contain appreciable admixture of $s\bar s$ pairs already 
at small momentum transfers. This paper 
discusses strangeness in the mean--field type of nucleon models,
and exemplifies this by explicit calculations in the MIT bag model
enriched by the presence of instantons. We calculate the instanton
contribution to the strangeness in the MIT bag (on top of the 
standard contribution to strangeness found in that model).
Although we do it in an essentially perturbative way, 
we present a detailed derivation of the formula expressing  
nucleon matrix elements of bilinear strange quark operators, 
in terms of a model valence nucleon state and interactions 
producing quark-antiquark fluctuations on top of that valence
state. We do it in detail to clarify our argument
that in the context of the mean--field 
type of quark models (where a Fock state expansion exists
and where the nucleon state can be constructed
out of single-quark states), the resulting formula 
acquires a significance beyond perturbation theory. 
The derivation combines the usage of the 
evolution operator containing a strangeness source,
and Feynman-Hellmann theorem. 
\end{abstract}



\newpage

\section{Introduction}

For quite some time already, a number of investigators has been
considering possibly nonzero strange matrix elements of non-strange
particles, nucleons \cite{DoN86,Ch76,Ch88,Er87,GaLLS88,BrEK88,BeJM88,
JaK87,PaR88,NoVZ89b,BlRS88,RhBP89,Ja89}. 
Such intriguing
discussions received additional impetus from the experimental \cite{As89}
and theoretical investigations of the related problem, namely
surprisingly small fraction of the nucleon spin carried by the quarks
\cite{RhBP89,Ja89,As88,JaM90,KaM88,HoM88,Ah87,Mc89,KlP89},\cite{BrEK88}. 
The intense 
research on nucleon strangeness continued into nineties up to 
the present day; references
\cite{IoK90,FrK90,KlP90,KlLVW90,PaSW91,VoW91,GaLS91,GaLS91b,PaW92,LuKW92,
ElFHK92,Do92,Mud92,CoFN93,MuD93,ElK93,DeNW93,Ad94,ElKKS95,DoL94,DoLL96,DoLW98,
We95,StW94,GeI97,KlKMP99} 
are just some of the examples. More precisely, matrix elements in 
question are $\langle N|{\cal O}_{s}|N\rangle$, where $|N\rangle$ 
is the nucleon state and ${\cal O}_{s}$ is an operator containing 
strange ($s$) quark fields bilinearly. In this section, the 
integration over the three-space is understood in the matrix element. 
Later, we will indicate the integration explicitly where needed. 
We will be concerned with ${\cal O}_{s} = \overline s \Gamma s$,
where $\Gamma$ is an arbitrary matrix in the spinor space.

Namely, although the {\it valence} component $|N_{0}\rangle$
of the full nucleon state $|N\rangle$ contains only $u$ and $d$ quarks,
quark-antiquark fluctuations 
include the $s\overline {s}$ component, allowing
\begin{equation}
  \langle N | \overline {s}\Gamma s |N\rangle \neq 0
\label{eq:sNumel}
\end{equation}
\noindent 
even though the net strangeness of the nucleon state $|N\rangle$ is of 
course zero. Some of these matrix elements may be surprisingly large, 
possibly pointing to some effects not expected in the naive quark model 
of hadrons.
{\it E.g.}, the strange scalar density inside the nucleon
is connected with the experimentally measured $\pi - N$ sigma-term
through the ratio

\begin{equation}
y = {\langle N| \overline {s} s | N\rangle\over {{\frac{1}{2}}
{\langle N| \bar u u + \overline d d |N\rangle}}}. \label{y-ratio}
\end{equation}

\noindent For example, see
\cite{DoN86}-\cite{Ja89},\cite{IoK90,GaLS91, GaLS91b,DeNW93}.
A review \cite{DeNW93} containing discussions of a very complete set of 
original references, estimates $y = 0.22 \pm 0.16$. 
Also, EMC experiment \cite{As88} provides evidence
that $\langle N| \bar s \gamma_\mu \gamma_5 s |N\rangle$
is possibly relatively large.

In the study of the long-debated issue of nucleon strangeness, the 
usage of nucleon models is still important. This of course rises 
the question of the model dependence --- even concerning the results 
on what is the basic mechanism behind the effect. For example,
an analysis of Steininger and Weise \cite{StW94} of the scalar 
strangeness of the nucleon
performed in the framework of the Nambu Jona-Lasinio (NJL) model,
obtained a very small
upper bound on the scalar strangeness from the NJL model with 
four-momentum cutoff, a larger but still modest upper bound on 
it from the NJL model with a three-momentum cutoff, but dramatically 
higher scalar strangeness arises when instanton-induced interaction 
among quarks dominates. 
In addition, these authors 
found only a small contribution, less than $3\%$, from kaon loops.
On the other hand, kaon loops are the basic mechanism 
for generating the nucleon strangeness in some other
approaches (see, {\it e.g.}, \cite{MuB94} and
Refs. in Sec. 2.1 in \cite{Mu94}, or discussion in
Forkel {\it et al.} \cite{FoNJC94}). 
Other examples are provided by the strangeness electric mean-square
radius, the sign of which is positive according to \cite{HoPM97,HoP93,GeI97},
but negative in some other approaches 
\cite{Ja89,PaSW91,MuB94,FoNJC94,Le96,Le95,HaMD96,Ch96}, 
or the strangeness nucleon magnetic form factor, for which 
predictions of various models and analyses range from +0.37 \cite{HoPM97,HoP93}
over positive \cite{Ja89,HaMD96} to various negative values 
\cite{Ja89,PaSW91,Mu94,Le96,Le95,HaMD96,Ch96} 
all the way down to possibly $-0.75\pm 0.30$ \cite{Le96,Le95}. 

This illustrates the motivation to investigate such issues further,
in as large number of different approaches as possible, 
attempting to decrypt what is the physics behind 
model dependence.
In the present paper, we formulate a framework which
will in principle enable comparison of 
such results
\cite{StW94} with corresponding results in a wider range 
of complementary models.
We also want to propose a framework which will
be applicable not only to
the scalar strangeness, but more generally. 
Below, we will give an expression for 
$\langle N |:\overline {s}\Gamma s:|N\rangle$ where 
$: ... :$ denotes normal ordering
with respect to the non-perturbative vacuum $|0\rangle$: 
\begin{equation}
   :\bar{q}\Gamma q: \: = \: \bar{q} \Gamma q - 
          \langle 0 | \bar{q} \Gamma  q | 0 \rangle \;.
\label{defNO}
\end{equation}
$\Gamma$ is an arbitrary matrix in the spinor space, say 
$\Gamma = 1_{4}, \gamma^{\mu}, \gamma_{5}, \gamma^{\mu} \gamma_{5}, 
\sigma^{\mu \nu},...,$ depending on whether one is interested in the 
scalar, pseudoscalar, vector, axial,
and for some purposes maybe even tensor, pseudotensor, {\it etc.}, 
... strangeness of the ``full" (model) nucleon state $|N\rangle$
which may contain $s\bar s$ pairs. Any interaction (call it  ${\cal
L}_I$) which can produce $s \overline{s}$ pairs 
can lead to such a nucleon state containing an intrinsic strangeness component.

\begin{figure}
\centerline{\includegraphics{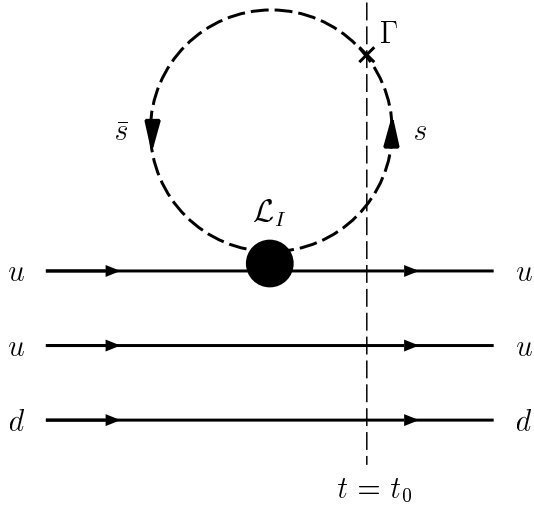}}
\caption{Non-vanishing nucleon strangeness due to 
a response of the valence nucleon state 
to a strangeness source at $\Gamma$ (denoted by $\times$), {\it i.e.} to 
a probe coupled to strange quarks through $\Gamma$.
More precisely, this graph is that part of the nucleon response which
arises only through one interaction ${\cal L}_I$.\label{probe}}
\end{figure}

That matrix elements $\langle N |\overline {s}\Gamma s|N\rangle$ 
can be significantly different from zero, is not very
surprising in nonperturbative QCD in the light of its non-vanishing quark 
scalar condensates -- 
the \emph{finite} vacuum expectation value\footnote{This is what is often 
-- {\it e.g.}, in O.P.E. -- denoted by $\langle 0| :{\bar q}q: |0\rangle$ 
$(q=u,d,s)$, but where the normal ordering is with respect to the 
{\it perturbative} vacuum, so that it does not vanish in the 
nonperturbative vacuum. We reserve the notation $: ... :$ for the
normal ordering with respect to the non-perturbative QCD vacuum.} 
of ${\bar s}s$ is 
actually approximately as large as for the non-strange quarks:
$\langle 0|\bar{s} s| 0\rangle \approx \langle 0|\bar{u} u| 0\rangle
= \langle 0|\bar{d} d| 0\rangle \approx (-240 {\rm MeV})^3$.
The MIT bag model provides a good illustration how this leads 
to a large $\langle N|\bar{s} s| N\rangle$ \cite{DoN86}. However,
there may also be $s\bar s$-pairs other than those from 
the QCD vacuum condensate, so that normal-ordered
strange operators can in principle also have non-vanishing 
nucleon matrix elements. Fig. \ref{probe}
illustrates how a non-vanishing value of not only 
$\langle N| \bar s \Gamma s |N\rangle$, but also 
$\langle N| :\bar s \Gamma s: |N\rangle$, can then 
get a contribution from these $s\bar s$-pairs 
not from the vacuum condensate: 
at the instant $t=t_0$ the composite nucleon is hit by an
external probe ({\it e.g.}, a neutrino \cite{Ah87}) 
with the coupling $\Gamma$ 
to the strange quarks. Due to an interaction capable of producing
$s\bar s$-fluctuations, the nucleon state $|N\rangle$
at the time-slice $t=t_0$ obviously contains not only the valence
quarks $uud$, but also the $s$--quark loop to which the
external probe can also couple. 
Let us {\it schematically} write down the full nucleon (proton) state
which is also coupled to the strangeness-sensitive probe:
\begin{equation}
 |N\rangle
     =  \frac {1}{\cal N} \left(
       \sum_{X=0}^\infty C_{X} |uud \, X \rangle +
       \sum_{X=0}^\infty C_{s\bar s X} |uud \, s\bar {s}\, X\rangle
\right)
 \equiv \frac {1}{\cal N} \Big(|N_{0}\rangle+| \delta N\rangle\Big)
\, , \label{fullproton}
\end{equation}
where $X$ (starting from $X=|0\rangle$ standing for the 
complicated non-per\-tur\-ba\-ti\-ve QCD vacuum) symbolizes any number of 
various perturbative {\it and} non-perturbative gluon configurations 
but also any number of
quark-antiquark pairs, including strange pairs which escaped
detection by this probe. 
These complicated configurations ``dress" quarks ($q=u,d,s ...$)
into their effective counterparts -- constituent quarks 
${\cal Q = U, D, S} \ldots$ .
(In terms of the constituent quarks, this part unperturbed 
by the strangeness-sensitive probe, is just the valence part: 
$| N_0 \rangle = |{\cal U U D} \rangle$. 
That $\langle N_0| :\bar s \Gamma s: |N_0\rangle = 0$
is especially obvious in terms of the constituent quarks.)  
The {\it one} strange pair detected at $\Gamma$,
has been explicitly denoted by $s\bar s$ in the
$|\delta N\rangle$-part of the nucleon state perturbed by the
probe. $|\delta N\rangle$ can be viewed as the response
of $|N_{0}\rangle$ to the weakly coupled strangeness-sensitive probe.
(The coefficients $C_X, C_{s\bar s X}$ denote the amplitudes of states
with various admixtures $X$ or ${s\bar s\, X}$.
${\cal N}$ is the normalization.) 
This response makes possible that in principle the total nucleon 
$\Gamma$--strangeness $\langle N| \bar s \Gamma s |N\rangle$ 
also receives a nonvanishing contribution from the non-vacuum 
channel $\langle N| :\bar s \Gamma s: |N\rangle$.

However, the question is how to get the nucleon state  
in specific enough terms in order to have a calculable expression 
for $\langle N |\!:\!\overline {s}\Gamma s\!:\!|N\rangle$.
To get it {\it exactly} would probably be tantamount to solving
nonperturbative QCD --- consider, for example, that the Fock
state expansion itself must be built upon the nonperturbative
QCD vacuum $(X=0)$, which is unknown. This is why we said that 
(\ref{fullproton}) is only a schematic, illustrative 
expression. Therefore, one obviously has to rely 
on models to a large extent. One seemingly more viable approach 
could for example be to {\it model} $|N_0\rangle$ in a conventional way
in terms of {\it only} non-strange effective quarks 
(so that $\langle N_0|\!:\!\overline s \Gamma s\!:\!|N_0\rangle=0$
even though  $\langle N_0 |\overline{s}\Gamma s|N_0\rangle \neq 0$
at least for $\Gamma = 1$ due to the strange vacuum condensate),
and then use appropriate interactions ${\cal L}_I$ to infect it 
by $s\bar s$-fluctuations and thereby produce $|\delta N\rangle$ -
say, using perturbation theory if ${\cal L}_I$ happens to be 
perturbative. This is what  (\ref{1stOrder}) in the next
section amounts to. However, in that section we also point out 
why one cannot proceed quite so straightforwardly, and then give 
our alternative formulation with the formula for matrix elements 
of strange operators. How this expression can be evaluated  
is explained in more detail in the third section, where 
we also explain why we are motivated to investigate the 
case of the instanton-induced  interaction. The evaluation
of various strange densities --- with this ${\cal L}_I$, and 
in a concrete nucleon model --- is carried out in the fourth 
section. We conclude in the fifth section. 

\section{Formulation of a model approach to nucleon  \\ stran\-ge\-ness}

As pointed out by Forkel {\it et al.} \cite{FoNJC94}, 
the (``naive'') absence of virtual $q\bar q$ pairs in the hadron 
wave functions in the models based on constituent quark core, 
makes the treatment of nucleon strangeness in such models
far from straightforward. Since the approach presented below 
is complementary to other ones which have also used 
dressed quarks in some way ({\it e.g.}, 
\cite{KaM88,StW94,FoNJC94}), we first 
give a review of some quark-model notions that will be
relevant below.

The purpose of working with hadron models is,
of course, not to solve but to {\it imitate} the horrendously
complicated non-per\-tur\-ba\-ti\-ve low-energy QCD. 
Accordingly, various gluon field configurations ({\it e.g.},  
instantons, or those configurations responsible for confinement) 
and polarization clouds of fluctuating  $q\bar q$ pairs 
(all symbolized by $X'$s in (\ref{fullproton})),
%that in principle exist in $|N_0\rangle$
as well as all interactions between all these fundamental constituents,
are taken into account through parameters of some nucleon model
and appropriate wave functions for dressed, {\it effective}
quarks and antiquarks $\cal Q=U,D,S$. 
Examples may be various constituent quark models, where 
baryons consist just of  
valence quarks which are however {\it constituent} quarks, 
quasi-particles which come about 
through dressing of the current quarks by QCD --- {\it i.e.}, 
in other words, by our $X$'s. Or, it may be the MIT bag model, where 
these long-range nonperturbative QCD effects lead to, or are 
partially parametrized by, a confining cavity
containing again a fixed number of effective valence quarks
(and antiquarks, in the case of mesons and 
"exotic" $qqq q\bar q$ baryons). 
Choosing a definite model of the hadron structure
%for an explicit computation of the nucleon matrix element of the
%strange operator of interest,
implies also the choice of the model wave function basis
$q_{K}(x)$ in which to 
expand the quark fields $q(x)$ ($q=u,d,s$) in terms of creation
(${\cal U}^\dagger_{K}, {\cal D}^\dagger_{K}, {\cal S}^\dagger_{K}$)
and annihilation (${\cal U}_{K}, {\cal D}_{K},{\cal S}_{K}$) 
operators of dressed quarks and antiquarks. ($K$ stands for the
set of quantum numbers labeling a model quark state. For the expansion
specific to the MIT bag-model see the Appendix.)
 
It is then clear, for example, that the nucleon $|\cal UUD\rangle$
(when all three of these effective quarks are in their ground states), 
is nothing but our $|N_0\rangle$ from (\ref{fullproton}) 
except that all the mess of fluctuations $X$ is by some model 
parametrization lumped into dressing of valence quarks $\cal UUD$,
as well as into effective model interactions, or a mean field they feel.
Obviously, the idea here is to represent hadrons as composed
of a fixed, well-defined number of dressed valence quarks (and
antiquarks), bound by effective model interactions which sum up 
reasonably successfully the fundamental QCD ones. 
%Although it is not necessary, let us for clarity simplify the 
%model picture a little bit further and view 
The most simplified, but illustrative case 
is when these model quasiparticles  
are moving in an average, mean field $\Phi$. 
%If these model interactions (and/or mean field)
%represent a satisfactory approximation to the real physical
%behaviour of the system, we say that the model is good, or
%at least that it is adequate for some purposes. 
To be sure, 
these model interactions (and/or mean field), as well as the 
%(parameters describing) 
effective, dressed quarks $\cal Q$, are assumed to be ``produced"
by all relevant interactions between quarks at more 
fundamental levels {\it including}
presently interesting strangeness-producing interactions ${\cal L}_I$. 
``Produced" here of course means that we {\it modeled}
them, not ``calculated" from these underlying relevant interactions.
%but in this low-brow, model way, 
So, they (including ${\cal L}_I$) are assumed to be accounted
for through modeling.
% as far as the dressing of the
%quarks is concerned, and as far as effectively summing up 
%their interactions in an average mean field is concerned.
 
Note, however, that this approach does not say what
would be the model representation (or parametrization)
of $|\delta N\rangle$, as it does for $|N_0\rangle
= |\cal UUD\rangle$.
Of course, in the spirit of all said above, we can write 
$|\delta N\rangle \sim |{\cal UUD S}\bar{\cal S}\rangle$, and
writing this is even quite useful for reminding us 
that {\it i)} the fluctuating strange (anti)quarks 
--- being embedded in the nucleon ---
also have to be dressed in the way prescribed
by whatever model is applied, including being
in one of the model single-quark eigenstates, and that 
%this 
modeling effectively takes care of all their interactions 
%they have with the non-strange sector of the bound state, 
(except of course the interactions induced by their coupling
at $\Gamma$ to their source, a probe sensitive to strangeness), 
and {\it ii)}, that all other fluctuations ($X'$s in
(\ref{fullproton})) are lumped in the dressing, so that the only 
allowed quark-antiquark fluctuation is ${\cal S}\bar{\cal S}$,
which has its source in the external strangeness-sensitive
probe at  $\Gamma$. 
However, in contradistinction to, {\it e.g.},  
$|N_0\rangle = |\cal UUD\rangle$ 
which is unambiguous because we know that there all quarks are in their 
model ground states (and corresponding quantum numbers on $\cal UUD$ are 
suppressed for brevity of the notation, but known in principle), 
$|\delta N\rangle \sim |{\cal UUD S}\bar{\cal S}\rangle$ is just
a generic formula, a useful mnemonic as just described, 
because in this case we {\it do not} know in what states
these five constituents are. In principle, $|\delta N\rangle$
is a superposition of all possible such states, encompassing
exotic baryons with ${\cal UUDS}\bar{\cal S}$ contents and ordinary 
strange baryons coexisting with kaons, as well as nucleons with 
$s\bar s$-mesons --- most notably $\phi$-mesons. 

So, let us call $%\hat
            {H}_{0}$ the Hamiltonian responsible for the
formation of hadron states composed of definite, fixed
numbers of quarks --- and possibly antiquarks.
In the simplest case, we can imagine $H_0$ as  consisting
of a sum of one-body quark operators, say 
typically of the effective quark kinetic energy operator $K$
and the mean, or self-consistent,  field $\Phi$ in which the 
dressed valence quarks would move. 
In any case, $H_0$ defines the nucleon model --- possibly together with
some other ingredients (like the confining boundary condition in bag 
models, for example).
The valence nucleon state $|N_{0}\rangle$ would then 
be the ground eigenstate, and $|k\rangle$ would stand for all 
possible higher eigenstates of ${H}_{0}$,
\begin{equation}
      {H}_{0}|N_{0}\rangle = E_{N_{0}}|N_{0}\rangle,
\qquad \qquad {H}_{0}|k\rangle = E_{k}|k\rangle,
\qquad \qquad  E_{k} >  E_{N_{0}}.
\label{Spectrum}
\end{equation}
For example, $ {H}_{0}$ could be the static bag model Hamiltonian. 
$|N_{0}\rangle$ would then be the bag model nucleon in its
ground state, and $|k\rangle$ all higher bag states with a definite
number of constituents, 
including also ``bagged" ${\cal UUD Q}\bar{\cal Q}$ exotic baryons
and the product meson-baryon bag states such as
$|k\rangle=\ket{\calU\calD\calQ}\ket{\calU\bar{\calQ}}$.

\begin{figure}
\centerline{\includegraphics{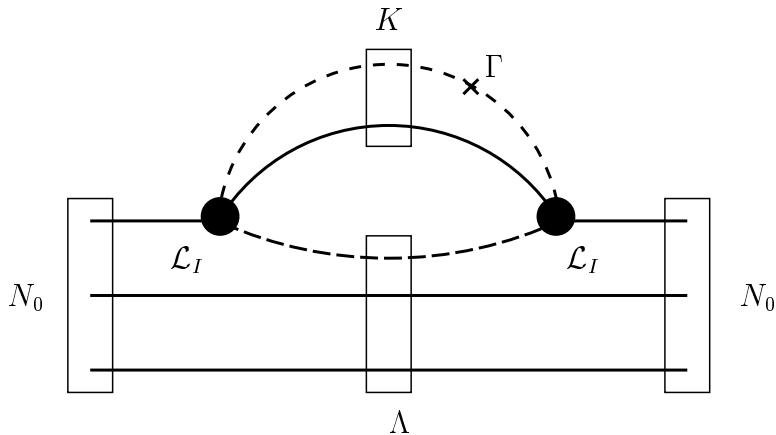}}
\caption{
A response of the valence nucleon state $|N_0\rangle$
to a strangeness source at $\Gamma$ through two interactions 
${\cal L}_I$. This type of contribution can be associated with 
the kaon-loop contribution to the nucleon strangeness
(a possible $K\Lambda$ intermediate state is therefore indicated).
\label{KLambda}}
\end{figure}

What $H_0$ cannot do is to produce $s\bar{s}$
fluctuating pairs. For that we have to invoke ${\cal L}_{I}$,
or its corresponding Hamiltonian ${H}_{I}$, as by assumption they 
can produce $s\bar{s}$
excitations on top of $|N_{0}\rangle$. 
To clarify that introducing ${\cal L}_{I}$ does not lead
to double-counting, let us repeat that $H_0$ is just a
model Hamiltonian, the parameters of which should mimic the
effects of full, true non-perturbative QCD as much as possible.
For example, if $H_0$ is the
Hamiltonian of the non-relativistic naive constituent quark
model, it must contain the postulated mass parameter of
the constituent quark mass $M_\mathcal{Q} \approx M_{N_0}/3$. The
corresponding quantity in the true theory, the dynamically
generated quark mass, is (in principle) the result of all
possible QCD interactions, so that the interactions related to
$H_I$ can, in real QCD, also contribute to this mass by contributing
to the $s\bar s$-fluctuations. The
dynamically generated non-strange quark mass must be close to
the model constituent quark mass parameter $M_\mathcal{Q}$ sitting
in
$H_0$, and only in such implicit, indirect ways are interactions
like $H_I$ ``present" in $H_0$. However, they are not present
explicitly, and, in fact, $H_0$ cannot produce any $s\bar s$
fluctuations at all. Therefore, if we want to study the $s\bar s$
fluctuations, we must introduce $H_I$ to enrich the model nucleon
with ${\cal S} {\bar{\cal S}}$-fluctuations on top of
$|N_0\rangle$.
Correspondingly, ${\cal L}_{I}$
(and thus also ${H}_{I}$)
contains strange quark field operators bilinearly so that it 
can connect $|N_{0}\rangle$ and  $|\delta N\rangle$ containing 
$s\bar{s}$ pairs. (This also implies
$\langle N_0 |:{H}_{I}:|N_{0}\rangle \equiv 
 \Delta^{(1)}E_{N}= 0$ regardless of what precisely 
this interaction is. This will be important  
in  (\ref{PerTh}) below, for the first-order shift 
 $\Delta^{(1)}E_{N}$ and the third-order shift  $\Delta^{(3)}E_{N}$.) 
In our figures, this interaction
is depicted as a two-body operator, where a strange quark bilinear
is combined with a non-strange bilinear. This may be, for example, the 
two-body part of ${\cal L}_{inst}$, the instanton-induced 
interaction\footnote{See, {\it e.g.}, ${\cal L}_{inst}$ of 
Shifman, Vainshtein
and Zakharov \cite{ShVZ80}, or its version used in \cite{Kl94}.}. 
%However, ${\cal L}_{I}$ can be more involved. 
On the other hand, Steininger and Weise \cite{StW94} 
studied the three-body part \cite{th76,th86} of instanton-induced interaction 
(which part they call ${\cal L}_{6}$). Nevertheless, the arguments 
here are completely general and encompass such cases too; one
would just have to do some obvious modifications in our figures.
(For example, in Figs. \ref{probe} and \ref{KLambda}, 
such a ${\cal L}_{I}$ would, in addition 
to the strange quark loop, straddle not just one but two valence
quark lines of different flavours.)  
So, $\langle N |\!:\!{\bar s}\Gamma s\!:\!|N\rangle$ could
be evaluated if $|\delta N\rangle$ could be found. But how?
For instance, it is easy to see that straightforward 
application of perturbation theory 
to find $|\delta N\rangle$,
where 
\begin{equation}
|\delta N\rangle =  \sum_{k \neq N_{0}}
\frac{\langle k | %\hat
{H}_{I}|N_{0}\rangle}
{E_{N_{0}}-E_{k}} \, |k \rangle + ...  \, ,
\label{1stOrder}
\end{equation}
is hardly viable even in those cases when ${H}_{I}$ would be truly
perturbative. Namely, it necessitates the summation over intermediate states
$|k\rangle$ (some of which must contain $s\bar{s}$-pairs,
in order to give $\langle N |\!:\!{\bar s}\Gamma s\!:\!|N\rangle\neq 0$),
which is very hard to handle in
practice. Admittedly, Geiger and Isgur \cite{GeI97} have
recently succeeded in performing such a straightforward perturbation
calculation of the proton strangeness (i.e., using  (\ref{1stOrder})).
However, in order to make their calculation tractable, they were 
forced to model hadrons
as simple harmonic oscillators. Also, the choice where to put a
cutoff, i.e. which intermediate hadron states $\ket{k}$ to discard,
is more ambiguous than when working with quarks.

Fortunately, the alternative formulation through the evolution operator 
is also possible. We formulated and used it in Ref. \cite{KlKMP99}.
Here we present a more detailed derivation of the expressions used
there. More concretely, we will combine 
the  Feynman-Hellmann theorem \cite{He37,Fe39}
with the usage of the evolution operator containing the 
Hamiltonian with the source of the strange current of interest. 
This is because in cases like this one, where we would like to 
avoid the need to construct $|\delta N\rangle$ explicitly, 
the method of sources is especially helpful. 
(See, {\it e.g.}, \cite{Ra81}, pages 89,90.) 
Of course, in this approach $|\delta N\rangle$ is the response
of $| N_0 \rangle$ to the external probe which is the source
of a strange current, and, naturally, we had this approach in mind 
already when we wrote schematically the full nucleon state coupled to 
a strangeness source as (\ref{fullproton}). 
This way we will not need $|\delta N\rangle$ explicitly.
Instead, we will obtain the nucleon matrix element of this current 
%$\overline {s}\Gamma s$  
as the response (to the current source) of the {\it transition 
amplitude} of the model ground state $| N_0 \rangle$ at 
$t\rightarrow -\infty$ into itself, but at $t\rightarrow +\infty$.
%to the strangeness source coupled to this strange quark current.

We will use normal-ordered operators in order to get an expression 
for $\langle N |\!:\!\overline {s}\Gamma s\!:\!|N\rangle$ which can
be non-vanishing due to the strange densities that may exist in the 
nucleon {\it on top} of the vacuum condensate densities that exist in 
the QCD vacuum. We are then in principle able to evaluate this matrix
element because we assume we 
can represent the ground state $| N_0 \rangle$ by a known nucleon 
model. Whatever we do below could be done also without normal ordering, 
but then the analogous expression would include the strangeness due 
to the strange condensate in the complicated QCD vacuum, and since we 
neither know the QCD vacuum state nor presently address its modeling, 
we cannot evaluate 
this expression. How to find the vacuum part of nucleon strangeness, is 
an issue that depends on the relation of each specific hadron model with
the QCD vacuum quark condensates. For example, this vacuum contribution
to the scalar strangeness was found in the MIT bag model by Donoghue
and Nappi \cite{DoN86}, while the expression for
$\langle N |\!:\!\overline {s}\Gamma s\!:\!|N\rangle$ 
evaluated in our Sec. 4, is the strangeness induced in the MIT 
bag nucleon valence ground state in addition to the vacuum
contribution.
 
Using the method of sources in combination with
the model approach will also enable us to use the perturbative
expansion of the evolution operator only formally; 
since there are plausible physical arguments,
which are {\it different} from the usual argument (used
in Ref. \cite{KlKMP99}) of ``smallness of
the perturbation", that higher orders should be neglected,
the resulting formula for strange nucleon matrix elements
should be applicable even when the interactions ${\cal L}_I$,
which lead to their non-vanishing values, 
is not really perturbative. These arguments are a novel
element with respect to Ref. \cite{KlKMP99}. 
Namely, as already emphasized, the
state used here as the ground state,  $| N_0 \rangle$, is fully
determined by some model Hamiltonian $H_0$
which also sums up the effects of ${\cal L}_I$ in dressing of the
constituent valence quarks, so that this interaction with strange 
quarks is not explicitly present --- just as the strange quarks are 
not explicitly present in $| N_0 \rangle$. However, ${\cal L}_I$ is again 
{\it induced} by the external strangeness source, because it brings in a
$s\bar s$ pair, which is not hidden in the modeled dressing polarization
clouds ($X$'s), and which must be absorbed via ${\cal L}_I$ by
the valence quarks so that the system is 
%ultimately 
returned to $| N_0 \rangle$. However, it turns out that a recognizable 
response within a defined nucleon model can be obtained only 
for a {\it limited} number of  ${\cal L}_I$-vertices, otherwise 
a double-counting occurs through dressing of already dressed quarks. 
All this will be explained more concretely below, after the
derivation, which we present in more detail than in Ref.  
\cite{KlKMP99} so that the common points as well as differences with
respect to Geiger and Isgur \cite{GeI97} are clear.

So,
%In order to profit from the Feynman-Hellmann theorem, 
let us define another, auxiliary perturbation Hamiltonian $%\hat
{H}^\prime$ by adding to $%\hat
{H}_{I}$ a source term
for the strange operator we want to calculate in the ``full"
nucleon state%$|N\rangle$
:
\begin{equation} 
%\hat
     {H}' \equiv %\hat
{H}_{I}+ \lambda \otimes \langle \overline{s} \Gamma s \rangle ,
\label{Hprime}
\end{equation}
where $\langle \overline{s} \Gamma s \rangle$ is the convenient abbreviation
\begin{equation}
\langle\overline{s} \Gamma s \rangle \equiv 
\int  \overline{s} (x) \Gamma s(x)\, d^{3}x \, . 
\label{sdens}
\end{equation}
(However, in all matrix elements $\langle N|\!:\!\overline{s} \Gamma s\!:\!|
N\rangle$ above and below, the three-space integration over 
$ \overline{s} (x) \Gamma s(x)$ is understood!)
The generic form $\lambda \otimes \Gamma$ can mean any of the cases 
$\lambda 1_{4}, \lambda_{\mu} \gamma^{\mu}, 
\lambda _{5 \mu} \gamma^{\mu} \gamma_{5}, 
\lambda_{\mu\nu} \sigma^{\mu \nu},...\,.$ 
%Lambdas $\lambda, \lambda_{\mu} , \lambda _{5 \mu} ,\, ... $ 
%are just auxiliary parameters, and when they are zero, 
%When the strengths of the external sources ( $\lambda, \lambda_{\mu} ,
%\lambda _{5 \mu} ,\, ... $) tend to zero, 
%the total Hamiltonian 
%$%\hat
%{H} (\lambda) = %\hat
%{H}_{0} + %\hat
%{H}'(\lambda)$ 
%containing the auxiliary sources, reduces to the physical one, 
%$%\hat
%{H} = %\hat
%{H}_{0} + %\hat
%{H}_{I}$. 
The Hamiltonians are normal ordered. It is usually implicitly understood, 
but, for clarity,  we will indicate normal ordering explicitly
everywhere in the remainder.

Obviously, the Feynman-Hellmann theorem applied to 
our $%\hat
	  {H}(\lambda)= {H}_{0} + {H}'(\lambda)$ 
says that the sought strange matrix element 
of the full 
%(and physical, at $\lambda =0$) 
nucleon state is 
\begin{equation}
\langle N| \int :\overline{s} (x) \Gamma s(x): d^{3} \! x |N\rangle = 
\left. \langle N | \frac{\partial :\! %\hat
{H} (\lambda)\! :}{\partial \lambda}|N\rangle \right|_{\lambda=0} =
\left. \frac{\partial E_{N} (\lambda)}{\partial \lambda}
\right|_{\lambda=0},
\label{derLamb}
\end{equation}
where we also took the physical limit of the vanishing strength 
($\lambda =0$) of the external strangeness source. 

Since $\Delta^{(1)}E_{N}=0$, the perturbed ground-state energy 
$E_{N}(\lambda)$ is 
%in the perturbation theory 
given by
\begin{displaymath}
E_{N} (\lambda) = E_{N_{0}} + 
%\langle N_{0} | %\hat
%:\! {H}' (\lambda)\! :|N_{0}\rangle 
 \sum_{k \neq N_{0}}
\frac
{\langle N_{0}|%\hat
:\! {H}' (\lambda)\! :|k\rangle \langle k|%\hat
:\! {H}' (\lambda)\! :|N_{0}\rangle}{E_{N_{0}}-E_{k}}
\end{displaymath}
\begin{displaymath}
+ \sum_{k,l \neq N_{0}}
\frac
{\langle N_{0}|%\hat
:\! {H}' (\lambda)\! :|k\rangle 
\langle k|:\! {H}' (\lambda)\! :|l\rangle 
\langle l|%\hat 
:\! {H}' (\lambda)\! :|N_{0}\rangle}
{(E_{N_{0}}-E_{k})(E_{N_{0}}-E_{l})}
+O[%\hat
        {H}' (\lambda)^{4}]
\end{displaymath}
\begin{equation}
\equiv       E_{N_{0}} + 
%\Delta^{(1)} E_{N} (\lambda) + 
	     \Delta^{(2)} E_{N}(\lambda)+ 
\Delta^{(3)} E_{N}(\lambda)+ \, O[%\hat
                           {H}' (\lambda)^{4}].        \qquad \qquad
\label{PerTh}
\end{equation}
%$|k\rangle$ are the eigenstates of the unperturbed Hamiltonian
%$ {H}_{0}$ as in  (\ref{Spectrum}). 
$O[{H}' (\lambda)^{4}]$ stands for
the fourth and higher orders, which, as we will argue soon, turn out not 
to contribute to %the nucleon strangeness 
(\ref{derLamb}). 
The presence of the sums over hadronic intermediate states
$\ket{k}$ of  (\ref{Spectrum}), similarly as in the approach of 
Geiger and Isgur \cite{GeI97} should be noted. One can then
easily understand how we can capture similar aspects of the
physics of nucleon strangeness in our respective approach.
Now, how do we expect to render the strangeness in (\ref{derLamb})
calculable, when (\ref{PerTh}) contains 
sums over intermediate states, and, as pointed out above,
handling them is precisely the 
difficulty that makes the conventional perturbative approach 
(as in    (\ref{1stOrder})) 
%and (\ref{NdelN})).
useless in practice? What 
%gets us off the hook 
helps here is that we can relate 
$\frac{\partial E_{N} (\lambda)}{\partial \lambda}$ 
to the nucleon matrix elements of the evolution operator 
$U(t_{2},t_{1})$, whose 
%conventional 
perturbation expansion is 
\begin{equation}
%\hat
{U} (t_{2},t_{1}) = 
1 + \sum^{\infty}_{n=1} %\hat
{U}^{(n)}(t_{2},t_{1})= \hat{T}
\Bigg\{ 1+ 
 \sum^{\infty}_{n=1}\frac{i^{n}}{n!}
\Big[\int^{t_2}_{t_{1}}\! :\! L_{\rm{int}}(t)\! : dt \, \Big]^{n} %+\, ...
\Bigg\}  \;.
\label{EvolOp}
\end{equation}
$\hat{T}$ denotes the time ordering operator and 
$L_{\rm{int}}(t)= \!\int \! {\cal L}_{\rm{int}} ({\bf{x}},t)d^{3}x 
= - H_{\rm{int}}(t)$ 
is the interaction Lagrangian. In our case, we should replace 
the interaction in the integrand with the form containing the 
strangeness sources, like in the definition of $%\hat
{H}^\prime$, 
 (\ref{Hprime}):
\begin{equation}
L{(t)}_{\rm{int}} \rightarrow L'(t) = 
L_I(t) - \lambda \otimes \langle \overline{s} \Gamma s (t)\rangle =
\int d^{3}x \Big[ {\cal L}_{I}(x)- \lambda \otimes 
\overline{s} (x) \Gamma s (x)\Big].
\label{Lprime}
\end{equation}
For definiteness, let us from now on
specialize $\lambda \otimes \Gamma$ to $\lambda_{\mu} \gamma^{\mu}$, 
{\it i.e.}, suppose that we are after the vector strangeness of the 
nucleon. It is trivial to re-formulate what follows for any other 
possible $\lambda \otimes \Gamma$. 
For example, the second order term in  (\ref{EvolOp}) 
is then 

\begin{eqnarray}
{U}^{(2)} (+ \infty, -\infty) &=& - \frac{1}{2}\, \hat{T} \int^{+
\infty}_{- \infty}\! dt \int^{+\infty}_{-\infty}\! dt' \, 
\Big[\, :\! L_{I}(t)\! :\, :\!  L_{I}(t')\! : \nonumber \\
 &-& \vphantom{\Bigg|}\lambda_{\alpha} :\langle\bar{s}
\gamma^{\alpha} s(t)\rangle: \,
:\!  L_{I}(t')\! : 
 -:\! L_{I}(t)\! : \, \lambda_{\beta} 
:\langle\bar{s} \gamma^{\beta} s (t') \rangle: \nonumber \\
&+& \vphantom{\Bigg|}\lambda_{\alpha} \lambda_{\beta} \,
:\langle\bar{s}\gamma^{\alpha} s(t)\rangle:\, :\langle\bar{s}
\gamma^{\beta}s(t')\rangle:\,\Big]  .
\label{U2}
\end{eqnarray}
 
The second- and third-order terms, ${U}^{(2)}$ and ${U}^{(3)}$, 
%, $%\hat
%{U}^{(2)}(+ \infty, - \infty)$ and ${U}^{(3)}(+ \infty, - \infty)$ 
are particularly interesting.
Their contribution to the $S$-matrix, when written with the help of the
interaction Hamiltonian $%\hat
{H}^\prime$ and the sum over intermediate states
$|k\rangle$, 
 
 \begin{eqnarray}
 S^{(2)}_{ab} &\equiv& \! \langle b | %\hat
{U}^{(2)}(+ \infty, - \infty)
       |a\rangle \! =\! - 2 \pi i \delta (E_{b}-E_{a})\! 
\vphantom{\Bigg|}\nonumber \\
&& \times \sum_{k \neq a}
	   \!  \frac{\langle b|\! :\! %\hat
{H}'(\lambda)\! :\!|k\rangle 
 \langle k |\! :\! %\hat
{H}' (\lambda)\! :\!|a\rangle}
 {E_{a}-E_{k}+i \epsilon} \, ,
\label{S2}
 \end{eqnarray}

 \begin{eqnarray}
 S^{(3)}_{ab} &\equiv & \! \langle b | %\hat 
{U}^{(3)}(+ \infty, - \infty)
       |a\rangle \! =\! - 2 \pi i \delta (E_{b}-E_{a}) \!
\vphantom{\Bigg|}\nonumber \\
 && \times \sum_{k,l \neq a}
         \!    \frac{\langle b|\! :\! %\hat
{H}'(\lambda)\! :\!|k\rangle  
\langle k |\! :\! %\hat
{H}' (\lambda)\! :\!|l\rangle
 \langle l |\! :\! %\hat
{H}' (\lambda)\! :\!|a\rangle} 
 {(E_{a}-E_{k}+i \epsilon)(E_{a}-E_{l}+i \epsilon)} \, ,
\label{S3}
 \end{eqnarray}

\noindent can obviously be related to $\Delta^{(2)}E_{N}$ and
$\Delta^{(3)}E_{N}$
%, the second- and third-order
%correction to $E_{N}(\lambda)$ 
in  (\ref{PerTh}) when $|a\rangle=|b\rangle=|N_{0}\rangle$:
 
 \begin{equation}
 \langle N_{0}|U^{(i)}{(+\infty, - \infty)}|N_{0}\rangle = 
    - 2 \pi i \delta (0) \Delta^{(i)} E_{N}(\lambda)\, ,\qquad i=2,3.
 \label{del2E}
 \end{equation}

\noindent [Strictly speaking, the divergence due to $\delta (0)$ renders
this expression meaningless; however, we will be able to 
get rid of $\delta(0)$.]
On the other hand, by using the standard 
field theory-expansion of $%\hat
{U}$, {\it i.e.},
 (\ref{EvolOp}), 
we avoid the need to consider the intermediate states $|k\rangle$.
To demonstrate this, let us for a moment concentrate on the
contribution to strangeness that comes from 
$U^{(2)}{(+\infty, - \infty)}$, given by (\ref{U2}):

\begin{eqnarray}
\lefteqn{\frac{\partial}{\partial \lambda_{\mu}} \langle N_{0} |U^{(2)} 
(+ \infty, - \infty)|N_{0}\rangle\Bigg|_{\lambda_{\mu}=0}
 =} && \hspace*{3cm}  \nonumber \\
 &=& \langle N_{0} | \frac{1}{2} \hat{T} \int^{+\infty}_{-\infty} dt
  \int^{+ \infty}_{-\infty}\! dt'\,
\Bigg[\, :\langle\overline{s} \gamma^{\mu} s (t)\rangle: \, :\!  L_{I}(t')\! : 
  \hspace*{3cm}\nonumber \\
&&\hspace*{5cm}+
:\! L_{I} (t)\! : \, :\langle\overline{s} \gamma^{\mu} s(t')\rangle:\,\Bigg]|N_{0}\rangle
\nonumber \\
 &=& \langle N_{0} |  \int^{+\infty}_{-\infty}\! dt\, \Bigg\{ \int^{t}_{-\infty} 
\! :\langle \overline{s} \gamma^{\mu} s (t)\rangle:\,\, :\!  L_{I}(t')\! :\,  dt' 
\hspace*{3cm} \nonumber \\
&&\hspace*{3cm} + \int^{+\infty}_{t} \!
:\! L_{I} (t')\! :\,\, :\langle\overline{s} \gamma^{\mu} s(t)\rangle: dt' \,
\Bigg\}
|N_{0}\rangle \;.
\label{eq:bezk}
\end{eqnarray}
Since 
%we have a static problem here, 
the nucleon strangeness cannot depend on the chosen time-slice $t$,
the expression in the curly brackets must be the same for 
any $t$. We can therefore fix $t$ in the curly brackets 
({\it i.e.} in the limits of integration and in 
$\langle\bar{s} \gamma^{\mu} s(t)\rangle$) to any constant value $t=t_{0}$
(say, $t_{0}=0$),
and we are free to factor out the expression in 
the curly brackets out of the integral over $t$:

\begin{eqnarray}
\lefteqn{\frac{\partial}{\partial \lambda_{\mu}} \langle N_{0} |U^{(2)}
(+ \infty, - \infty)|N_{0}\rangle\Bigg|_{\lambda_{\mu}=0} = 
\left(\int^{\infty}_{-\infty} dt\right) \langle N_{0} | \Bigg\{ }&&
\nonumber \\
&\times&\!\!\!\!\!\! \int^{t_0}_{- \infty} 
\!\!\! :\langle\overline{s} \gamma^{\mu} s(t_0)\rangle:\, \,
:\!  L_{I}(t')\! :\, dt' 
+\! \int^{+ \infty}_{t_0}  \!\!\! :\! L_{I} (t')\! :\,\,
:\langle \overline{s} \gamma^{\mu} s(t_0)\rangle: dt'
\,\,\Bigg\} |N_{0}\rangle \;.
\nonumber \\
\label{factorOut}
\end{eqnarray}

\noindent This then leaves the integral 
$(\int^{+\infty}_{- \infty}dt)$ 
as a constant but divergent prefactor. 
However, it exactly matches the constant divergent prefactor in 
 (\ref{del2E}),
$2 \pi \delta (0)= \int^{\infty}_{-\infty} dt$, 
and they cancel each other out.
%This means that we are done, 
The inspection of the  (\ref{factorOut}), (\ref{del2E}), (\ref{PerTh})
and (\ref{derLamb}) 
then gives the contribution of $U^{(2)}$ to the nucleon strangeness. 
This is the first term on the right-hand side of  (\ref{formula})
below (where we have again gathered the time-ordered integrals
into one from $-\infty$ to $+\infty$ but containing the time-ordering 
operator $\hat T$).
Repeating the above procedure for $U^{(3)}$ gives us 
the second term in  (\ref{formula}). {\it I.e.}, the strange 
nucleon matrix element of the full nucleon state is then given by 

%\vspace{-1mm}
\begin{eqnarray}
\lefteqn{\langle N|:\overline{s} \Gamma s:|N\rangle \, 
%\left. \frac{\partial\Delta^{(2)}E_{N} (\lambda)}{\partial \lambda}
%\right|_{\lambda=0} +
%\left. \frac{\partial\Delta^{(3)}E_{N} (\lambda)}{\partial \lambda}
%\right|_{\lambda=0} 
= \, i \, %\{ 
\int^{+ \infty}_{- \infty}\! dt' \,
\langle N_{0}|{\hat T} :\langle\overline{s}\Gamma s (t_0)\rangle:\, 
:\! L_{I} (t')\! : |N_{0}\rangle }\hspace*{1cm} &&
\nonumber \\
 &\hspace*{-2cm}-& \hspace*{-1cm} \frac{1}{2}
 \int^{+\infty}_{- \infty}\! dt' \,  \int^{+\infty}_{- \infty}\! dt'' \, 
\langle N_{0}|{\hat T}:\langle\overline{s} \Gamma s(t_0) \rangle:\,:\! L_{I}(t')\! :\, 
 :\! L_{I}(t'')\! :|N_{0}\rangle      %\}
\label{formula}
\end{eqnarray}

\noindent (We have reverted from the special case of $\gamma_\mu$
to a general matrix $\Gamma$.) 

%Consistently with earlier definitions, 
%$<\overline{s} \Gamma s>$ and $L_{I}$
%are the three-space integrals over the densities  
%$\overline{s} (x) \Gamma s(x)$ and ${\cal L}_{I}(x)$, respectively.

Obviously, the non-vanishing contributions to (\ref{formula})
occur only when the strange quark fields are fully contracted. 
{\it E.g.}, the integrand of the first term  in
(\ref{formula}), written in terms of space integrals
over the strange current and Lagrangian densities is
\begin{eqnarray}
\lefteqn{\int \! d^{3}\! x  \,  d^{3}\! x' \langle N_{0}|{\hat T} 
 :\overline{s} (x)
\Gamma s (x): \, :{\cal L}_{I} (x'): |N_{0}\rangle}
\hspace*{2cm} &&\nonumber \\
&=& \int \! d^{3}\! x \, d^{3}\! x' \langle N_{0}| :\overbrace{\overline{s}
(x) \Gamma \underbrace{s(x) {\cal L}_{I}}}
(x'): |N_{0}\rangle \;,  \label{eq:contractions}
\end{eqnarray}

\begin{figure}[htb]
\centerline{\includegraphics{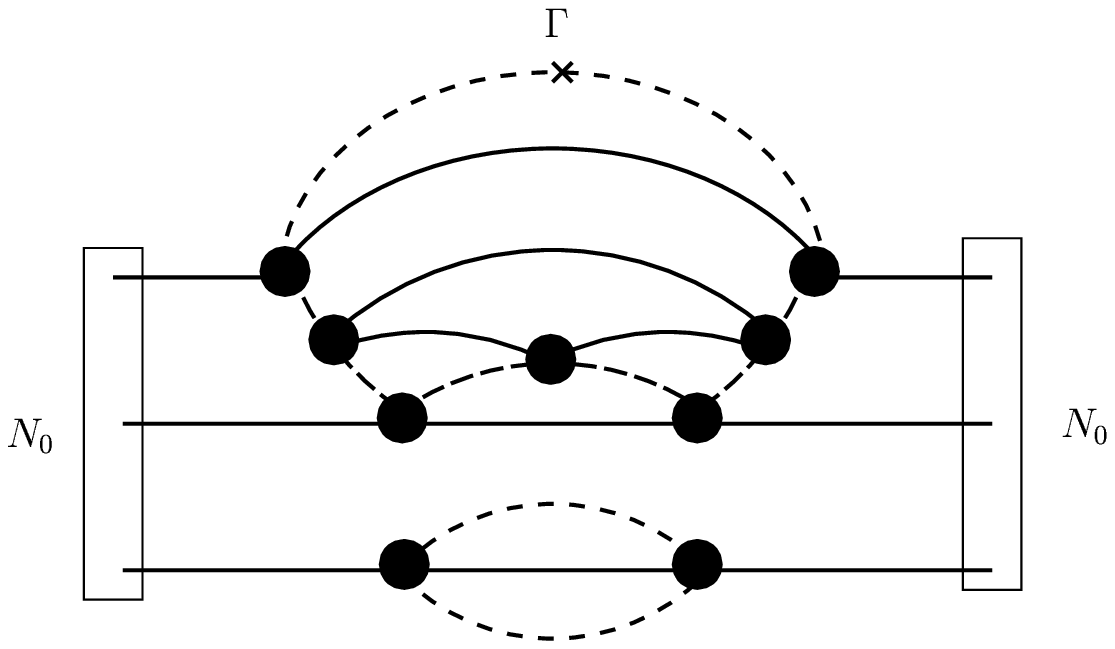}}
\caption{
One of the spurious graphs that would originate from
the tenth-order term $U^{(10)}$ of the evolution operator.
It illustrates how the contributions from the terms higher
than $U^{(3)}$ cannot be identified with responses to a
strangeness source at $\Gamma$, but instead lead to dressing of
already
dressed quarks and to changing of the defined model interactions.
As in the previous figures, black dots denote the interactions
${\cal L}_I$, solid lines denote nonstrange dressed quarks, and
dashed lines strange dressed quarks.
\label{u10}}
\end{figure}

\noindent (where the contractions are indicated by over- and
underbraces, and $t_0 \equiv x_0, t'\equiv {x'}_0$ for consistency
of the notation). So, the first term in (\ref{formula}) corresponds
to Fig. \ref{probe}, since these contractions, or time-ordered pairings, 
are of course
the propagators of strange quarks. In the second term, the two contractions
must connect the strangeness source at $\Gamma$ with two different 
separately normal-ordered interaction Lagrangian densities
which act as ``sinks" for strangeness at two different points of
a valence quark line, or two different  valence quark lines.
In any case, there must be an additional strange quark contraction between 
these two $:{\cal L}_I:$'s, and this completes the  strange quark loop.
Fig. \ref{KLambda} gives an example of the graphs originating from the second
term of (\ref{formula}), namely the $U^{(3)}$ contribution. 
Clearly, this way kaon loops can be generated. If the result
of \cite{StW94} on small contribution of kaon loops is not 
an artifact of their model, it is likely that the second term 
in (\ref{formula}) will be much smaller than the first one if
(\ref{formula}) is evaluated in realistic enough models. However, this
cannot be known in advance. So, why not include still higher 
contributions which would give contributions like Fig. \ref{u10} for example? 

First, let us remember that there are some 
perturbative interactions ${\cal L}_I$  which can
nevertheless be important for nucleon strangeness.
One example is the depletion of the instanton density
which occurs \cite{Kl94} in the MIT bag model, and which is so strong
that in this case one can for sure treat the instanton-induced 
interactions perturbatively, using the density of very dilute instantons
as the expansion parameter. In the light of results of \cite{StW94}
which indicate that instanton-induced interactions may be an
especially important source of strangeness, even such a perturbative
contribution from instanton-induced interactions may well be
significant.

So, in some cases we can justify 
ignoring  the contributions of $U^{(n)}, n \ge 4$,
({\it i.e}, $O[{H}' (\lambda)^{4}]$ in  (\ref{PerTh}))
as a perturbative approximation\footnote{Had we limited 
ourselves to such cases, we could have 
omitted most of our discussion of the role of
dressed quarks in Secs. 1 and 2, because we would not
have needed such detailed explanations for justification
of the formula (\ref{formula}).}.
However, we are actually better off than that, because our model
prediction for strangeness terminates with the $U^{(3)}$
contribution even for nonperturbative interactions ${\cal L}_I$,
because it turns out that contributions from higher $U^{(n)}$'s
would be double-counting. 
This follows from our view on nucleon strangeness 
as the response (to a strangeness-sensitive probe)
of nucleon model states which, in the static regime 
(before or after any interactions with external probes),
are just $|N_0\rangle$, {\it i.e.}, are by assumption built 
only of non-strange {\it dressed} quarks ${\cal U,D}$,
which hide all the complexity of QCD -- including strange 
fluctuating pairs -- in their (modeled) dressing.
Remember that
except the interactions induced by the strangeness source,
all fundamental interactions (including ${\cal L}_I$) and 
resulting fluctuations of gluon {\it and} quark fields are
lumped in forming these effective quarks and their effective
model interactions and/or mean field they feel. Now, the contribution
to (\ref{formula}) from a $\lambda$-differentiated term 
$U^{(n)}, n\geq 4$ (through  (\ref{eq:bezk})) would correspond to graphs
with one vertex at $\Gamma$ from which two propagators of dressed
strange quarks would emanate, two ${\cal L}_I$-vertices which
would receive them (so it would be like in Fig. \ref{KLambda}, from the
$\lambda$-differentiated $U^{(3)}$
contribution, so far), but then other $n-3$ ${\cal L}_I$-vertices would
follow. ({\it E.g.}, see Fig. \ref{u10}.)
Depending on how the contractions are arranged, they can
be connected in the loop originating at $\Gamma$, 
or can be disconnected from it, forming their separate loops.  
%coming from chained contractions of strange fields with and 
%among remaining $:{\cal L}_I:$ in $U^{(n)}$. 
In both cases this is obviously double counting, as 
these additional $n-3$ vertices represent dressing of 
%already dressed 
quarks that have already been dressed. 
The first two ${\cal L}_I$-vertices are different,
as they are induced by the strangeness source --- they are
the unavoidable sink for the $s\bar s$ pair created by the source 
at $\Gamma$. The second term in (\ref{formula}), {\it i.e.}, the
$U^{(3)}$ contribution, is the highest possible term that has just that, 
and does not contain
additional interactions of the dressed strange loop with already
dressed valence quarks, resulting in double-dressing.  
%An equivalent way of looking at it is saying that 
%$|\delta N\rangle \sim |{\cal UUD  S}\bar{\cal S}\rangle$
%is the response of $|N_0\rangle = |{\cal UUD}\rangle$
%to the strangeness-sensitive probe. This ``strange response
%quarks" are sucked out of valence ${\cal U,U,D}$ quarks 
%via ${\cal L}_I$ interactions, but at most two of them 
%because the probe coupled at $\Gamma$ can receive just two
%strange quark lines, and closing the strange loop through 
%additional ${\cal L}_I$-interactions with valence quarks ....
%{\cal S}\bar{\cal S}$
A completely equivalent, but probably even clearer way 
to see this is to view our external strangeness-sensitive
probe as a {\it sink} (instead of a source) of a strange quark 
current. So, when this sink (by means of interaction ${\cal L}_I$)
sucks a strange pair out of the polarization clouds that form our 
dressed valence quarks, this strange pair should go to this
sink at $\Gamma$ that pulled it out ---  and {\it not}
%(before ending up at $\Gamma$)
%first 
run all around the nucleon interacting 
%via ${\cal L}_I$ 
with the valence quarks up to $n-3$ times ($n=4, 5, ... , \infty$), 
%before ending up at $\Gamma$
also altering in the process already defined model interactions 
between the valence quarks. 

Realizing this also automatically answers why there is no
such contractions among the {\it non}strange quark fields
which would lead to additional ${\cal U}\bar {\cal U}$ and
${\cal D}\bar {\cal D}$ pairs. Such loops would also
appear if contributions from higher $U^{(n)}$ could 
enter in  (\ref{formula}). (See Fig. \ref{u10}.) {\it I.e.}, the avoidance
of double-dressing gives the response of $|N_0\rangle$
to a strangeness source in the generic form 
$|\delta N\rangle \sim  |{\cal UUD S}\bar{\cal S}\rangle$ and not 
$|{\cal UUD S}\bar{\cal S}{\cal S}\bar{\cal S}{\cal D}\bar {\cal D}...\rangle$,
{\it etc.}, without imposing by hand any additional limitations to 
``one-particle, one-hole" responses. 

\section{Strangeness evaluation with a specified interaction ${\cal L}_{I}$}

Evaluation of the ``master formula" (\ref{formula}) is in principle
straightforward once one specifies two things: {\it i)} the overall
description of the hadronic structure, which amounts to choosing
the mean-field Hamiltonian $H_0$ in (\ref{Spectrum}), and {\it ii)}
${\cal L}_{I}$, which generates the $q\bar q$ fluctuations.
Namely, specifying {\it i)} should normally define also the single quark
solutions; a concrete calculation within a specified framework
or a model involves
expanding of quark fields in an appropriate wave function basis
({\it e.g.}, in the next section we choose to employ the quark 
solutions for the MIT bag). The field contractions in (\ref{formula}) 
lead to the sums over stationary modes of single quarks
and antiquarks, or, equivalently, the bound state propagators of 
these dressed model quarks. These sums require a regularization,
but now, for the single quark modes, it is much easier to physically
justify the choice of the cutoff than in the case of the sum over the
exotic baryon and meson-baryon states like the one in 
 (\ref{1stOrder}) or in ref. \cite{GeI97}. Let us recall
at this point that in the course of our derivation 
(\ref{Hprime})--(\ref{eq:contractions}), we replaced
these summation over hadronic states by summation over the
states/modes of quarks which constitute these hadrons. The
sum over quark modes should naturally run only up to some typical
hadronic low-energy cutoff $\Lambda \sim$ 0.6 GeV -- 1 GeV. This cutoff
on quark energies is dictated by
the fact that nonperturbative interactions among quarks operate
at low energies, whereas they gradually weaken and go over to the
perturbative regime for higher energies. 
In the aforementioned study of ${s\bar s}$ effects of
kaon loops \cite{GeI97}, Geiger and Isgur have shown the importance 
of high-mass intermediate states in these loops.  However, since 
these are hadronic, meson--baryon intermediate states, this does not 
contradict with cut-off such as $\Lambda \sim 1$ GeV on {\it quark}
energies. Namely, the dominant portions of the results of Ref. 
\cite{GeI97} are accounted for by states lying below 3--3.5 GeV. 
For comparison, our cut-off of 1.1 GeV (see Table 1) imposed 
on the energies of one strange {\em quark} and one {\em antiquark} 
fluctuating on top of the valence nucleon state, corresponds to
total energies up to $2 \Lambda + M_N\sim$ 3 GeV as well.
This leads us to believe that we have accounted for the majority 
of important degrees of freedom in a way compatible with
Ref. \cite{GeI97}.

The cutoff values of 0.6 GeV -- 1 GeV are 
typical for calculations in models of low-energy QCD, {\it e.g.},
the Nambu and Jona-Lasinio (NJL) model \cite{StW94}.
Obviously, we are supposing
here that the nucleon strangeness is the effect of low-energy,
nonperturbative QCD. Indeed, this brings us to the point
{\it ii)}, {\it i.e.} to the question what to use
concretely for ${\cal L}_{I}$ in  (\ref{formula}) in the 
explicit calculation -- performed in the next section -- 
of $\langle N|:\overline {s}\Gamma s:|N\rangle$. 

${\cal L}_{I}$ can of course be any interaction which can produce
fluctuating $s \overline{s}$ pairs, but
the question is, which interactions can be important in producing the
strangeness of the nucleon? For example, perturbative QCD
interactions
%one-gluon exchange
should be relatively unimportant in this regard.
Although precisely the
perturbative, high energy deep inelastic scattering reveals the sea of
$q \overline{q}$ pairs, including $s \overline{s}$,
%and this
%perturbative $q\bar q$ (and gluon) sea is even rather well understood, since
%it behaves in agreement with predictions of perturbative QCD.
%However,
%the share of the momentum carried by the strange perturbative sea has
%been measured by deep inelastic scattering and is now known to be relatively
%unimportant\cite{Oltman,Voss,IoK90,28}.
the contribution of this perturbative sea to the nucleon
strange matrix elements has traditionally been judged as relatively
unimportant --- see, {\it e.g.} Refs. \cite{IoK90,DeNW93}. A theoretical analysis
\cite{JiT95} of the CCFR data \cite{Ba95} on strange quark
distribution functions from neutrino-nucleon deep inelastic scattering,
seems to further support this point of view. For example, it finds a
very small upper bound on the strange radius of the nucleon
$(|\langle r^2 \rangle_s| \leq 0.005 $ fm$^2$) \cite{JiT95}
when extracted from such
parton distribution functions characterizing the nucleon structure at
high momentum transfers.
The  possibly enhanced nucleon strangeness
%of some other operators
is thus expected (see {\it e.g.} \cite{IoK90})
as an effect of nonperturbative QCD which, at low energies, around
nucleon mass scale, is certainly more important for hadronic structure than
perturbative QCD, and can lead to $s\bar s$ pairs already at small
momentum transfers, {\it i.e.}, large distances.
Nonperturbative QCD is after all responsible for
precisely such effects as forming quark-antiquark condensate $\langle 0|
\bar{q} q | 0\rangle$ $(q = u, d, s)$ and gluon condensate characterizing
the nonperturbative QCD vacuum. Some investigators (see, {\it e.g.},
ref. \cite{GeI80,DiP84,DiP86}, or, for a recent and 
comprehensive review,
ref. \cite{Sh95}) have suggested that among the most important nonperturbative
configurations of the gluon fields
are instantons. By now it is certainly well-established that the effective
interaction between quarks resulting from the presence of instantons (let us
call this interaction ${\cal L}_{inst}$), plays a very important role in the
formation of hadron structure \cite{Sh95} although it is not responsible for
confinement \cite{Gr85,Si89}, as thought previously. (In the present approach,
confinement must anyway be taken care of by the unperturbed
Hamiltonian $H_0$.) This ${\cal L}_{inst}$ is therefore in our opinion 
worth testing as an important candidate for the
interactions ${\cal L}_I$ generating the strange nucleon matrix
elements of some operators. 
%Indeed, there are indications that
%${\cal L}_{inst}$ is the most important part of ${\cal L}_I$:
A recent calculation \cite{StW94} in the context of NJL model 
%corroborates this point of view, 
seems to be an indication that ${\cal L}_{inst}$ is indeed
the most important part of ${\cal L}_I$,
as it found large strange pair components in the nucleon 
only if instanton-induced interaction was included in the 
low-energy dynamics. In that case, the ratio $y$ (\ref{y-ratio})
can be several times larger than its
upper limit in the case when the ``standard'' NJL model is used,
even  when augmented by kaon cloud effects \cite{StW94}.

Here we quote the vacuum-averaged version of the instanton-induced
interaction ${\cal L}_{inst}$ derived by ref. \cite{NoVZ89} in the instanton
liquid approach but transformed to $x$-space.
It is actually convenient to
separate it in one-, two-, and three-body pieces, ${\cal L}_{1},
{\cal L}_{2}$ and ${\cal L}_{3}$ respectively:

\begin{equation}
{\cal L}_{inst} = {\cal L}_{1} + {\cal L}_{2} + {\cal L}_{3},
\label{Linst}
\end{equation}

\begin{equation}
{\cal L}_{1} =  -n\left(\begin{array}{c}
{4\pi { } ^{2}\over 3}{\rho ^{3}}\end{array}\right)
\Bigg\{{\cal F}_{u}\, \bar{u}_{R}\, u_{L}
+ (u\leftarrow\joinrel \rightarrow d)
+ (u\leftarrow\joinrel \rightarrow s) \Bigg\}
+ (R \leftarrow\joinrel \rightarrow L) 
\label{eq:L1}
\end{equation}

\begin{displaymath}
{\cal L}_{2} =  -n\left(\begin{array}{c}
{4\pi { } ^{2}\over 3}{\rho ^{3}}\end{array}\right) ^{2}
\Bigg\{ {\cal F}_{u}\, {\cal F}_{d}\,
\Big[(\bar{u}_{R}u_{L})(\bar{d}_{R}d_{L})+ {3\over 32}
\big(\bar{u}_{R}\lambda ^{a}u_{L}\bar{d}_{R}\lambda ^{a}d_{L}
\qquad\qquad
\end{displaymath}
\begin{equation}
 - {3\over 4} \bar{u}_{R}\sigma _{\mu \nu }\lambda ^{a}u_{L}
\bar{d}_{L}\sigma ^{\mu \nu }\lambda ^{a}d_{L}\big)\Big]
+ (u\leftarrow\joinrel \rightarrow s)
+ (d\leftarrow\joinrel \rightarrow s) \Bigg\}
+ (R\leftarrow\joinrel \rightarrow L)
\label{eq:L2}
\end{equation}

\begin{displaymath}
{\cal L}_{3} =  -n\left(\begin{array}{c}
{4\pi { } ^{2}\over 3}{\rho ^{3}}\end{array}\right) ^{3}
\, {\cal F}_{u}\, {\cal F}_{d}\, {\cal F}_{s}\, {1\over 3!}\, {1\over
{N_{c}(N^{
2}_{c} - 1)}} \,
\epsilon_{f_{1} f_{2} f_3} \, \epsilon _{g_{1} g_2
g_3}\Bigg\{ \frac{2 N_c +1}{2 N_c +4}
\end{displaymath}
\begin{equation}
\!\!\times (\bar{q}^{f_{1}}_{R}q^{g_{1}}_{L})
(\bar{q}^{f_{2}}_{R}q^{g_{2}}_{L})(\bar{q}^{f_{3}}_{R}q^{g_{3}}_{L}) +
{8\over 3(N_{c}+3)} (\bar{q}^{f_{1}}_{R}q^{g_{1}}_{L})
(\bar{q}^{f_{2}}_{R}\sigma_{\mu \nu }q^{g_{2}}_{L})
(\bar{q}^{f_{3}}_{R}\sigma^{\mu \nu }q^{g_{3}}_{L}) \Bigg\}
\label{eq:L3}
\end{equation}

\noindent Here, $n$ is the instanton density and ${\cal F}_{f}$'s
are the characteristic
factors (corresponding to inverse effective quark masses) composed of
current light quark masses $m_{f}$ $(f=u,d,s)$,
average instanton size $\rho \simeq \frac{1}{3}$ fm \cite{Sh82,DiP84,DiP86},
and the quark condensate
$\langle 0|\overline{q} q|0\rangle = (-240 {\rm MeV})^{3}$.
{\it E.g.}, for the $u$-flavour, ${\cal F}_{u} \equiv
[m_{u}\rho - \frac{2\pi^{2}}{3} \rho^{3} \langle 0| \overline{q}
q|0\rangle ]^{-1}$, and analogously for the other flavours.
The left (and right) projected components are
defined in the usual way; {\it e.g.}, for the $u$-flavour,
%\begin{equation}
$u_{L,R} = \gamma_{\pm} u \equiv \frac{1}{2} (1 \pm \gamma_{5}) u$.
%\end{equation}

\begin{figure}
\centerline{\includegraphics{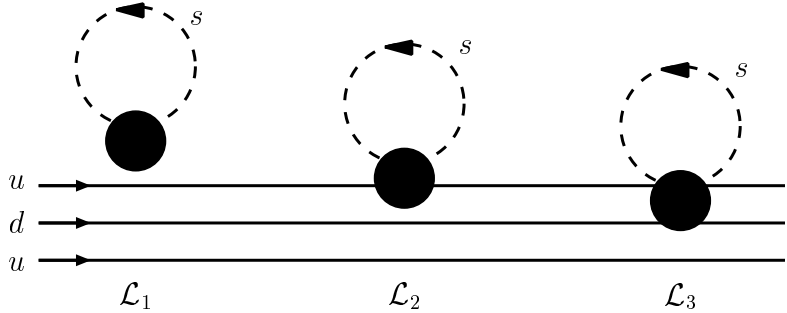}}
\caption{Instanton-induced local strangeness represented by the
effective one-, two- and three-body operators. Non-strange quarks are
denoted by solid lines, and strange ones by dashed lines.\label{L123}}
\end{figure}

In the three body interaction ${\cal L}_{3}$, the indices $f_i$, $g_i$
$(i=1,2,3)$ run over light flavours $u,d,$ and $s$.
{\it E.g.}, $g_{3}=d$ means $q^{g_3}_{L} = d_{L}$.
Repeated indices are summed over. The interaction defined here by
${\cal L}_{1}, {\cal L}_{2}$, and ${\cal L}_{3}$ is actually the
same as the well-known  one of Shifman, Vainshtein and Zakharov
(SVZ) \cite{ShVZ80},
although the present three-body term (\ref{eq:L3}) looks much simpler, but it
is just that Nowak \cite{No91} Fierzed away otherwise very complicated color
structures in that piece of SVZ interaction \cite{ShVZ80}, reshuffling them
in simple prefactors involving the number of quark colors $N_{c}$.
Obviously, the two-body term is the one which, through the 
(\ref{formula}) and (\ref{eq:contractions}), yields the graph in
Fig. \ref{probe}.
In addition to that, there is also a contribution to the nucleon
strangeness due to the three-body interaction ${\cal L}_{3}$,
exemplified by the last loop in Fig. \ref{L123}.
Such graphs come about when contractions in (\ref{eq:contractions})
are done with a strange  bilinear in ${\cal L}_{3}$.

In contradistinction to ${\cal L}_{2}$ and ${\cal L}_{3}$,
the contribution to the nucleon
stran\-ge\-ness due to the one-body term ${\cal L}_{1}$ does not involve
any interacting with the valence quarks, as illustrated in Fig. 3.
Perhaps not surprisingly, this disconnected graph requires some care.
${\cal L}_{1}$ has in fact the form of a mass term, and can be thought
of as the self-energy, or the effective mass that a quark acquires from
the effective interaction caused by the instanton liquid through which
quarks move in the nonperturbative QCD vacuum. Now imagine that we want
to evaluate some strange nucleon matrix element (\ref{formula}) in
some kind of constituent quark model where one  from the start uses
effective constituent quark masses already ``dressed" by nonperturbative
QCD. The self-mass part of the instanton effects would in that case
be already included in the constituent mass parameters. Using ${\cal
L}_{1}$ in the calculation would therefore be double-counting, so
in that case it must be dropped. On the other hand, if we would use
some approach where one uses the current, Lagrangian quark masses,
like in the MIT bag model for example,
there is no reason to drop ${\cal L}_{1}$ and it should be included in
the calculation on equal footing with ${\cal L}_{2}$ and ${\cal L}_{3}$.

We also note that the average instanton size
$\rho \simeq \frac{1}{3}$ fm $=(600 {\rm MeV})^{-1}$
is consistent with what we said above about the typical
hadronic cutoff scale $\Lambda \sim $ 0.6--1 GeV.
Namely, the effective interaction ${\cal L}_{inst}$ cannot
be operative at energies which would probe distances
significantly smaller than the average size of these
extended objects, instantons, which produce ${\cal L}_{inst}$.
%the extended objects that induce this
%interaction among the quarks which simultaneously
%``see" such an extended object.
%extended objects

The  final point we should clarify concerns consistency of using the
instanton induced interaction ${\cal L}_{inst}$ for ${\cal L}_{I}$ in 
(\ref{formula}), even in the case when we view  (\ref{formula}) as
a purely perturbative result.

Namely, although in the previous section we have advanced the arguments 
why the applicability of  (\ref{formula}) goes beyond perturbation
theory in the context of some quark models, we want to
point out that even if we forget for a moment these arguments, 
what we do in the next section can be justified
already from a viewpoint that is essentially perturbative. 
So, if we take this viewpoint, why is  (\ref{formula})
applicable not only to parts of ${\cal L}_{I}$ which come from
perturbative interactions like the perturbative gluon exchange, but
also to ${\cal L}_{inst}$   (\ref{Linst})--(\ref{eq:L3})
which is of nonperturbative origin, precisely wherefrom its
importance derives in this low-energy context? The point is that the
{\it origin} of ${\cal L}_{inst}$ is nonperturbative,
{\it i.e.}, these effective interactions
between quarks are the consequence of nonperturbative gluon configurations
 --- instantons. However, ${\cal L}_{inst}$  itself contains a small
parameter,
namely the instanton density $n$, and it is in fact so small that perturbative
expansion in its powers is possible.
Original estimates \cite{Sh82} where
$n \approx 1.6 \cdot 10^{-3}$ GeV$^{4}$ proved to be quite reliable as they
 have remained essentially unchanged \cite{Sh95} also
in the more recent instanton liquid calculations.
It is in fact useful to
define a ``dimensionless instanton density" $\tilde n$ by expressing it in
units of the average instanton size $\rho$, $n \equiv {\widetilde n}
\rho^{-4}$. The commonly accepted value is $\rho = 1/600$ MeV$^{-1}
\simeq 1/3$ fm \cite{DiP84,DiP86,Sh88}.
Therefore $\widetilde{n} \simeq 12.4 \cdot 10^{-3} \simeq 1/81$
and this dimensionless parameter
is obviously small enough to be used as the parameter of the perturbative
expansion. (The expansion parameter in QED is not much smaller,
$\alpha \simeq 1/137$.)
We should also keep in mind that this is the instanton density
in the true, nonperturbative QCD vacuum, while in some circumstances the
appropriate $n$ can be even smaller. Notably, 
ref. \cite{Kl94} found that in the MIT bag model
enlarged with the instanton--induced interaction 
(\ref{Linst})--(\ref{eq:L3}), 
which is used in the next section 
for the first evaluations of the  nucleon strangeness
via formula (\ref{formula}), the instanton density is very
strongly depleted with respect to the true QCD vacuum. 
Ref. \cite{Kl94} used certain approximations and assumptions, 
so that the depletion may be not quite so strong as estimated there,
but the usage of ${\cal L}_{inst}$ in (\ref{formula}) is clearly
consistent anyway, since even the aforementioned value of the
undepleted instanton density in the truly nonperturbative
QCD vacuum is small enough to serve as the parameter of the perturbative
expansion.

%\vspace{-1mm}
\section{Instanton--induced strangeness in the MIT bag model}
%\vspace{-1mm}

Now, we turn to the actual calculation of strange nucleon matrix elements
in the MIT bag model and with the instanton-induced interaction
$\mathcal{L}_{\rm inst}$ given by  (\ref{eq:L1}-\ref{eq:L3}).
For definiteness, we
quote the results for the proton---since the neutron case is quite similar,
we keep  $\ket{N}$ (for nucleons) in our expressions.

Using  (\ref{formula}), the proton-strangeness matrix element is
\begin{eqnarray}
\langle N | : \bar{s} \Gamma s : | N \rangle &=& i \int^{\infty}_{-\infty}
dt' \, \langle N_0 |\, \hat{T} : \int d^3x\: \bar{s}( \mbox{\boldmath$x$},
t_0) \Gamma s ( \mbox{\boldmath$x$}, t_0):\nonumber \\
 &\times& :\int d^3 y\: \mathcal{L}_{\rm inst}
( \mbox{\boldmath$y$}, t'): | N_0 \rangle \;,
\end{eqnarray}
where we have kept only the first term in the perturbation series
over low instanton density.  We have treated each of the three
parts of $\mathcal{L}_{\rm inst}$ (\ref{Linst}) separately. The one-body
interaction $\mathcal{L}_1$ (\ref{eq:L1}) is the simplest of all. Since no
valence quarks take part in this interaction, the only relevant part
of $\mathcal{L}_1$ is
\begin{equation}
-n \left( \frac{4\pi}{3}\rho^3\right) \mathcal{F}_s (\bar{s}_R s_L
 + \bar{s}_L s_R) \;,
\end{equation}
giving the $\mathcal{L}_1$ part of the matrix element:
\begin{eqnarray}
\langle N | : \bar{s} \Gamma s : | N \rangle_{\mathcal{L}_1}
 &=& i \int^{\infty}_{-\infty} \langle N_0 | N_0 \rangle \:
\hat{T} : \int d^3x\: \bar{s}( \mbox{\boldmath$x$},
t_0) \Gamma s ( \mbox{\boldmath$x$}, t_0):\nonumber \\
&\times& :\int d^3 y\: \bar{s}( \mbox{\boldmath$y$},t')
s( \mbox{\boldmath$y$}, t'): \;.
\end{eqnarray}
By taking into account the expansion of the strange quark
quantum field $s(x)$ in the MIT bag-model wave functions $s_K(
\mbox{\boldmath$x$})$ (see Appendix), this reduces to the
\begin{eqnarray}
\lefteqn{\langle N | : \bar{s} \Gamma s : | N \rangle_{\mathcal{L}_1}
\;=\; 4 \pi^2 n \rho^3 \mathcal{F}_s \sum_{K,L} \frac{1}{\omega_K +
\omega_L}} \nonumber \\
&\times& \left\{ \int d^3 x\: \bar{s}_M ( \mbox{\boldmath$x$}) \Gamma
s_{N}^{c} (\mbox{\boldmath$x$}) \int d^3 y\: \bar{s}_{N}^{c}
( \mbox{\boldmath$y$}) s_M ( \mbox{\boldmath$y$})\; + \; (s \leftrightarrow
s^c) \right\}\;.
\label{sgs}
\end{eqnarray}
Here, $K$ and $L$ stand for sets of quantum numbers labelling quark
states in the bag $K=\{n, \kappa, j_3\}$, $L=\{n', \kappa', j'_3\}$
(see Appendix). The sum over $K$ and $L$ goes up to the state with
$n=1$, $\kappa=-1$ (corresponding to the cut-off of about 1.1 GeV),
encompassing four lowest-lying strange quark states displayed in
Table \ref{tab1}.

\begin{table}
\begin{center}
\begin{tabular}{|c|c|c|}
\hline
$n$ & $\kappa$ & $\omega_{n\kappa}$ /MeV \\ \hline
0 & -1 & 514.0\\
0 & -2 & 726.7\\
0 & 1 & 797.4 \\
1 & -1 & 1104.9 \\ \hline
\end{tabular}
\caption{\small Strange quark states which can be excited by the
instanton interaction.\label{tab1}}
\end{center}
\end{table}

The expression for the contribution of the two-body interaction
$\mathcal{L}_2$
is somewhat more complicated,
involving also valence quark wave functions. Luckily, the terms with
$\sigma^{\mu\nu}$ cancel out, leaving us with
\begin{eqnarray}
\lefteqn{\langle N | : \bar{s}\Gamma s : | N \rangle_{\mathcal{L}_2}
\;=\; \frac{16}{3} \pi^4 n \rho^6 \mathcal{F}_q \mathcal{F}_s \sum_{K,L,\pm}
\frac{1}{\omega_K + \omega_L}} \nonumber \\
&&\times\left\{\int d^3 x\: \bar{s}_K ( \mbox{\boldmath$x$})\Gamma
s^{c}_L ( \mbox{\boldmath$x$})\;\int d^3 y\: \bar{s}^{c}_L
( \mbox{\boldmath$y$})\gamma_{\pm} s_{K}( \mbox{\boldmath$y$})
\vphantom{\Bigg|}\right. \nonumber \\
&&\hspace*{1cm} \times \Bigg[
2 \bar{q}_{0,-1,\frac{1}{2}}( \mbox{\boldmath$y$})\,\gamma_{\pm}\,
q_{0,-1,\frac{1}{2}}( \mbox{\boldmath$y$})\; + \;
\bar{q}_{0,-1,-\frac{1}{2}}
( \mbox{\boldmath$y$})\:\gamma_{\pm}\:q_{0,-1,-\frac{1}{2}}
( \mbox{\boldmath$y$})\Bigg]  \nonumber \\
&&+ \int d^3 x\: \bar{s}^{c}_K ( \mbox{\boldmath$x$})\Gamma
s_L ( \mbox{\boldmath$x$})\;\int d^3 y\: \bar{s}_L
( \mbox{\boldmath$y$})\gamma_{\pm} s^{c}_{K}( \mbox{\boldmath$y$})
\nonumber \\
&&\hspace*{1cm} \left. \times \Bigg[
2 \bar{q}_{0,-1,\frac{1}{2}}( \mbox{\boldmath$y$})\,\gamma_{\pm}\,
q_{0,-1,\frac{1}{2}}( \mbox{\boldmath$y$})\; + \;
\bar{q}_{0,-1,-\frac{1}{2}}
( \mbox{\boldmath$y$})\:\gamma_{\pm}\:q_{0,-1,-\frac{1}{2}}
( \mbox{\boldmath$y$})\Bigg] \right\}  \;.
\end{eqnarray}
Here $q_{0,-1,\pm\frac{1}{2}}(\mbox{\boldmath$y$})$ is the wave function
for the ground state of the  valence quark in the bag, which we take to be
the
same for $u$ and $d$ quarks.

Going now to the three-body interaction $\mathcal{L}_3$, expressions
become extremely long and complicated, so we do not write them down
here. In any case, as seen below,  it turns out that this contribution
is much smaller than the preceding two.

After focusing on the
scalar ($\bar{s}s$) and pseudoscalar ($\bar{s}\gamma_5s$)
stran\-ge\-ness as the channels
preferred by the QCD-vacuum fluctuations \cite{Zh97} we have
checked the vector ($\bar{s}\gamma_{\mu}s$) and the axial-vector
($\bar{s}\gamma_{\mu}\gamma_5s$) channels too.

The calculation of the contribution of the two-body, ${\mathcal{L}_2}$,
and three-body, ${\mathcal{L}_3}$,
instanton interactions is tedious and
in manipulation
of all these formulae we have relied heavily on \emph{Mathematica}
package \cite{Wo88} for symbolic computer calculations.

To get a rough idea how the calculation in the MIT bag model was
performed and in which way such a model choice influences our results, we
briefly sketch the calculation with the one-body part
${\mathcal{L}_1}$ interaction bellow.

\subsection{Scalar and pseudoscalar strangeness}

Let us first consider the \textbf{scalar} strange current density
$\bar{s}s$ inside the proton. The
expression for the matrix element can be written as
\begin{equation}
\langle N(p')|\bar{s}s| N(p) \rangle = A_{s}(q^2)
\bar{u}_N(p') u_N(p) \;,
\label{defsc}
\end{equation}
where $q^2 = (p-p')^2$, and $u_N$'s are nucleon
spinors. $A_{s}(q^2)$ is the scalar form factor accounting at
$q^2 =0$ for the scalar strangeness of the proton.

Calculations inside the bag model can be performed by making
the substitution $\Gamma =1$
and inserting the appropriate quark and antiquark wave functions
in (\ref{sgs}).
By a simple calculation one can show that the surviving combination
is the one with $\kappa = -1$, $\kappa' = 1$ and $\kappa =1$,
$\kappa' = -1$. Therefore,
\begin{eqnarray}
\lefteqn{\langle N | : \bar{s}s : | N \rangle_{\mathcal{L}_1}
\;=\; 4 \pi^2 n \rho^3 \mathcal{F}_s
\sum_{K,L,{\kappa,\kappa' = -1,1}}\hspace*{-8.5mm}\rule{0pt}{1ex}^{'}
\hspace{1cm}\frac{1}{\omega_K +
\omega_L}} \nonumber \\
&\times& \left\{ \int d^3 x\: \bar{s}_K ( \mbox{\boldmath$x$})
s_{L}^{c} (\mbox{\boldmath$x$}) \int d^3 y\: \bar{s}_{L}^{c}
( \mbox{\boldmath$y$}) s_K ( \mbox{\boldmath$y$})\; +
\; (s \leftrightarrow s^c) \right\}\;,
\end{eqnarray}
where ${\sum}'$ denotes the incomplete sum where the cases with equal
$\kappa$ quantum numbers are omitted, and
\begin{eqnarray}
\lefteqn{\langle N | : \bar{s}s : | N \rangle_{\mathcal{L}_1}
\;=\; 4 \pi^2 n \rho^3 \mathcal{F}_s \sum_{n=0}^{1} 4}\nonumber \\
&\times&\! \bigg[  2 N_{-1}(x_{n,-1}) N_1(x_{0,1})
\int r^2 dr W_{+}(n,-1) W_{-}(0,1) j_0(x_{n,-1}\frac{r}{R})
j_0(x_{0,1}\frac{r}{R}) \nonumber \\
&& + W_{-}(n,-1) W_{+}(0,1) j_1(x_{n,-1}\frac{r}{R})
j_1(x_{0,1}\frac{r}{R}) \, \bigg]^2 \;.
\end{eqnarray}

The normalizations $N_{\pm 1}(x_{n,\pm 1})$ and the
$W_{\pm}$-factors are given in the Appendix.
The above equation represents the contribution to the strange
scalar form factor
$A_{s}(q^2 = 0)$ coming from the one-body interaction. The remaining
contributions from the ${\mathcal{L}_2}$ and ${\mathcal{L}_3}$ instanton
interactions can be calculated similarly and the results are
\begin{eqnarray}
\langle N | : \bar{s}s : | N \rangle_{\mathcal{L}_1} &=&0.035\;, \\
\langle N | : \bar{s}s : | N \rangle_{\mathcal{L}_2} &=&0.023\;, \\
\langle N | : \bar{s}s : | N \rangle_{\mathcal{L}_3} &=&
2.9\cdot 10^{-4} \;.
\end{eqnarray}
Summing them up gives
\begin{equation}
A_{s}(0)_{\mathcal{L}_{\rm inst}} = 0.058.
\label{037}
\end{equation}
The evaluation of space-integrals was performed numerically, using the
following values for the parameters: 
the bag radius $R$=1/197.3 MeV$^{-1}\approx$1 fm,
the average instanton size $\rho$=1/600 MeV$^{-1}$, and the instanton
density $n=2.66 \cdot 10^7$ MeV$^4$, which is depleted instanton
density in the MIT bag as found in \cite{Kl94}.
Moreover, we take the
strange quark mass $m_s$=200 MeV and the valence
quark mass $m_u=m_d\equiv m_q$=8 MeV. The quark condensate
that follows from the Gell-Mann--Oakes--Renner relation
for these quark masses and the empirical meson masses is
$\langle 0| \bar{q}q|0\rangle\approx (-200 \textrm{MeV})^3$.

The \textbf{pseudoscalar} strange form factor $B_{s}$ is defined  as 
\begin{equation}
\langle N(p') | \bar{s} \gamma_5 s e^{-i \vec{q}\cdot
 \vec{x}}| N(p) \rangle = 
B_{s}(q^2) \bar{u}_N(p') \gamma_5 u_N(p) \;.
\end{equation}
For the pseudoscalar strange current $\bar{s}\gamma_5 s$,
 (\ref{sgs}) gives the vanishing one-body contribution
\begin{equation}
\langle N | : \bar{s} \gamma_5 s : | N \rangle_{\mathcal{L}_{1}}
= 0 \;.
\end{equation}
Analogously, we  obtain the vanishing result for 
the other two instanton interactions, i.e. 
$\langle N | : \bar{s} \gamma_5 s : | N \rangle_{\mathcal{L}_{\rm inst}}
= 0 \;$.
We thus obtain 
\begin{equation}
B_{s}(0)_{\mathcal{L}_{\rm inst}} = 0 \;,
\end{equation}
as the vanishing total instanton contribution to the pseudoscalar
form factor.

\subsection{Vector and axial-vector strangeness}

Recently, there has been a lot of experimental activity 
\cite{Mu97,An99} devoted to the \textbf{vector} strangeness, described
by Dirac ($F_{1}$) and Pauli ($F_{2}$) form factors 
in the decomposition
\begin{equation}
\bra{N}\bar{s}\gamma_{\mu}s\ket{N}=\bar{u}_{N}(p')\left[
F_{1}^{s}(q^2)\gamma_{\mu} + F_{2}^{s}(q^2)
\frac{i\sigma_{\mu\nu} q^{\nu}} {2 M_{N}}\right]u_{N}(p) \;.
\label{f1f2}
\end{equation}

For the comparison with the experimental data, the Sachs form factors,
$G_E$ (electric) and $G_M$ (magnetic) are widely used:
\begin{eqnarray}
G_E(q^2) &=& F_1(q^2) + \frac{q^2}{4 M_N^2} F_2(q^2) \, ,\nonumber\\
G_M(q^2) &=& F_1(q^2) + F_2(q^2) \, ,
\end{eqnarray}
with definitions $e G_E^{(0)} = Q$ (charge) and 
$(e/2 M_N) G_M^{(0)} = \mu$
(magnetic moment). By taking the non-relativistic nucleon spinor
\begin{equation}
u_N(p,s) = \sqrt{\frac{E+M_N}{2 E}} \left(
\begin{array}{c}
\chi_s \\
\frac{\displaystyle{\vec{\sigma}\cdot\vec{p}}}{
\displaystyle{E+m}} \chi_s
\end{array}
\right)
\end{equation}
in the Breit frame, defined by 
\begin{eqnarray}
q^{\mu} &=& (q^{0}, \vec{q}) = (0,\vec{q}_B) \;, \nonumber\\
\vec{p} &=& \frac{\vec{q}_{B}}{2}\; , 
\; \vec{p'} = -\frac{\vec{q}_{B}}{2} \, ,
\end{eqnarray}
the components of the vector current receive the form
\begin{eqnarray}
\langle N(p',s')| V_0^s | N(p,s)\rangle &=& \frac{m}{E}
\chi^{\dagger}_{s'}\chi_s G_E^{(s)}(-q^2_B) \;, \nonumber\\
\langle N(p',s')| \vec{V}^s | N(p,s)\rangle &=& \frac{1}{2 E}
\chi_{s'}^{\dagger} i (\vec{\sigma}\times \vec{q}_B) \chi_s 
G_M^{(s)}(-q^2_B) \;.
\label{defGm}
\end{eqnarray}

In order to calculate the contribution from the instanton induced vector strange
current inside the MIT bag, we have to identify the form factors in
(\ref{defGm}) with the Fourier transformed vector current within the bag
\begin{equation}
\langle N(p')|:\! V_{\mu}^{s}(q^2)\!:| N(p)
\rangle_{\mathcal{L}_{\rm inst}} =
\langle N(p')|: \!\! \int d^3\vec{r} e^{-i\vec{q}_B \cdot \vec{r}} 
\bar{s}(\vec{r})
\gamma_{\mu} s(\vec{r}): |N(p)\rangle_{\mathcal{L}_{\rm inst}},
\end{equation}
in the static limit $q \rightarrow 0$. Simple check with $V^s_0(q^2)$
component of the vector current gives zero, {\it i.e.} 
$G_E^{(s)}(q^2=0)_{\rm inst}= 0$ as it should be.

A similar calculation for the space components $\vec{V}^s$ shows
a non-trivial cancellation among the contributions of quarks in the
loop with different spin orientations, producing the total result
\begin{equation}
G_M^s(0)_{\mathcal{L}_{\rm inst}} = 0 \, .
\label{vector0}
\end{equation}
This implies the vanishing strange magnetic moment
\begin{equation}
\mu_{s}=F_{2}^{s}(0) = 0 \;,
\label{mus}
\end{equation}
which is compatible with the recent measurements at MIT/Bates 
\cite{Mu97} and even more recent ones at TJNAF (JLab) \cite{An99}.

The estimation of the \textbf{axial-vector} strangeness can be done along the 
same lines. The form-factor 
decomposition, assuming the $G$-parity 
symmetry of strong interactions, has the form
\begin{eqnarray}
\lefteqn{\langle N(p')|\bar{s}\gamma_{\mu}\gamma_5 s|N(p)\rangle  }
\nonumber \\
&=& \bar{u}_N(p')
\left( \gamma_{\mu}\gamma_5 G_1^{s}(q^2) + \frac{q_{\mu}}
{2 M_N}\gamma_5 G_{2}^{s}(q^2)\right) \bar{u}_N(p) \;.
\end{eqnarray}

The instanton contribution to such a matrix element can be calculated as
\begin{eqnarray}
\lefteqn{\langle N(p')|: A_{\mu}^{s}:| N(p) 
\rangle_{\mathcal{L}_{\rm inst}} }\nonumber \\ && =
\langle N(p')|: \int d^3 r e^{-i\vec{q}_B\cdot
\vec{r}} \bar{s}(\vec{r})
\gamma_{\mu} \gamma_5 s(\vec{r}):|N(p)
\rangle_{\mathcal{L}_{\rm inst}}
\end{eqnarray}
and should be compared with the axial form factors defined in the Breit 
frame as 
\begin{equation}
\langle N(p',s')|\vec{A}^{s}| N(p,s) \rangle = G_{A}^{s}(0) 
\chi_{s'}^{\dagger} \mbox{\boldmath$\sigma$} \chi_s \;.
\end{equation}
Again, it turns out that the axial-vector strangeness induced by the instanton
interaction is vanishing,
\begin{equation}
G_{A}^{s}(0)_{\mathcal{L}_{\rm inst}} = 0 \; .
\end{equation}

\section{Discussion and conclusions}

The original MIT bag model \cite{Ch74,Ch74b,DeG75} represents a 
suitable starting point in predicting the low-energy properties 
of low-mass hadrons. In this model, $R_{\rm bag}$ corresponds to the 
separations $R_{\rm confining}\sim$ 1 fm at which confinement 
effects are important, arising at the scale 
$\Lambda_{QCD}\simeq$ 100 to 300 MeV. Short--distance effects are 
supposedly taken care of by the perturbative one--gluon exchange.

However, in order to account for the effects at intermediate distances, 
{\it i.e.} at the momentum scales $Q \sim \Lambda_{\chi SB}\simeq$ 0.6 
GeV, the effective interaction (\ref{Linst})-(\ref{eq:L3}), 
induced by the liquid of small instantons 
(of the average size $\rho = 1/3$ fm) appears appropriate. 
Of course, the effects of the instanton--induced interactions are not 
included in Donoghue and Nappi's \cite{DoN86} simple bag-model relation
\begin{equation}
\bra{N}\bar{s}s\ket{N} = - \bra{0}\bar{s}s\ket{0} V
\label{naive}
\end{equation}
for the scalar nucleon strangeness, and the relative
importance of this naive strangeness and the instanton effects
is precisely what interests us here.

An advantage of the formula (\ref{formula}) is that in
principle it treats
the scalar, pseudoscalar, vector, axial, tensor or pseudotensor
nucleon stran\-ge\-ness in a unified manner; one just has to specify
what $\Gamma$ is. Within a chosen nucleon model, the evaluation
of (\ref{formula}) would proceed in --- essentially --- the same
way for each $\Gamma$, except for technical details.
Nevertheless, these technical details make a huge difference
in practice because, as clarified in the previous section,
even if the scalar and pseudoscalar cases are tractable,
the tensor or pseudotensor cases seem prohibitively hard
to do. However, this is a significant difference only now, at
the present capabilities of symbolic manipulation software,
and will diminish with the certain advancement of this
software and computer power in the future.

In the scalar case ($\Gamma=1$), the naive bag-model strangeness 
(\ref{naive}) is actually rather large for standard values of parameters. 
For our values, given at the end of subsection 4.1, it is
\begin{equation}
A_s^{\rm Nbag} \equiv -\bra{0}\bar{q}q\ket{0} V_{\rm bag}=4.36 \;,
\end{equation}
which is much larger than the instanton-induced contribution
(\ref{037}), and dominates the summed strangeness
\begin{equation}
A_s \equiv A_s^{\rm Nbag} + A_s(0)_{\calL_{\rm inst}} = 4.42 \;.
\end{equation}

Owing to using a somewhat smaller value of the quark condensate, Do\-no\-ghue
and Nappi \cite{DoN86} obtained 3.6 for this naive strangeness,
which is still rather large. $A_s^{\rm Nbag}$ depends very
strongly on the model size parameter $R_{\rm bag}$ since
$V_{\rm bag}=R_{\rm bag}^3 4\pi/3$. For example, $A_s^{\rm Nbag}$ would
decrease by a factor of 2 if $R_{\rm bag}=0.8$ fm, a nucleon size
which may be more acceptable, as the standard MIT bag value of
1 fm seems too large ({\it e.g.}, see \cite{BrKRW88}).
However, since the model dependence on the bag radius is similar for
other presently interesting matrix elements, the model dependence
largely cancels out when one forms ratios.
In particular, the instanton-induced contribution (\ref{037})
remains small in comparison with the naive nucleon bag strangeness,

\begin{equation}
\frac{A_s^{\rm Nbag}}{A_s(0)_{\calL_{\rm inst}}} \sim 75 \;,
\end{equation}
for reasonable variations of the radius parameter.

Note that using ${\cal L}_{inst}$ for ${\cal L}_I$ in (\ref{formula})
enables one to see what happens in different models with the intriguing
results of Steininger and Weise \cite{StW94} concerning the importance 
of the instanton-induced interaction for the scalar strangeness of nucleons.
Our results in the MIT bag model happen to disagree with their results in 
the NJL model enlarged with 't Hooft's instanton-induced interaction. Our 
results indicate that the instanton-induced interaction contributes just 
a small fraction to the -- otherwise rather large \cite{DoN86} --
scalar strangeness of nucleons modeled as MIT bags. 

Obviously, the contribution due to the difference in the
condensate with respect to the true, non-perturbative QCD
vacuum dominates the stran\-ge\-ness in the nucleon bag.
Admittedly, the instanton-induced contribution of this size
{\it would} be obtained in the calculation of  (\ref{037}) 
\emph{if} one would --- inside the MIT bag --- use the non-depleted
instanton density $n=1.6\cdot 10^9$ MeV$^4$. However, we consider
this merely as a consistency check, and not as an
alternative description of strangeness in the MIT bag, because using
the instanton density appropriate to the non-perturbative
QCD vacuum containing the large quark condensate,
would imply assuming the nonperturbative QCD
vacuum and the quark condensate not only outside, but also
inside the bag. This would indeed enable $A_s(0)_{\calL_{\rm inst}}$
to replace $A_s^{\rm Nbag}$ in full, but would also make the
MIT bag description inconsistent \cite{Kl94}.

The diluteness of the instanton liquid justifies 
the one-instanton approximation ({\it i.e.}, the first 
order in the perturbation theory for ${\cal L}_{inst}$)
indicated in Fig. \ref{L123}. 
The second--order contributions to  (\ref{formula})
should be even smaller than 
the small first--order results on instanton--induced strangeness 
we obtained in the MIT bag model. This 
removes the motivation for evaluating them, at least 
in the framework of that model.
Of course, in some other models, and possibly also with some
other ${\cal L}_I$, the results on the nucleon strangeness
it induces can be considerably higher, making the evaluation
of the second--order contributions more interesting.
As commented above, if one would find in different
models that the second term in (\ref{formula}) is
small in comparison with the first term in (\ref{formula}),
one would corroborate the result of \cite{StW94}
that virtual kaon loops
contribute little to the scalar strangeness.
For the reasons explained above, this conclusion
indeed seems natural in the present approach. 
More generally, our result (\ref{formula}) may well
help to clarify the relationship (which, {\it e.g.},
\cite{Mu94} judges as rather unclear) between
the kaon loop contribution, and the $\phi$-meson pole
contribution \cite{Ja89}, or the vector-meson ($\phi,\omega$)
dominance contribution \cite{FoNJC94}. Namely,
we believe that it will be possible (for $\Gamma = \gamma_\mu$)
to relate the first term in (\ref{formula}) to such
$\phi$-meson contributions in a way similar to the
relationship of the second term in (\ref{formula})
with the kaon loop contribution.
More recent evaluation \cite{FoN99} based on the up-to-dated
information on the (soft) nucleon-hyperon-$K^{\ast}$ form
factors yields the results reduced by more than an order
of magnitude. This brings the vector strangeness closer
to our result (\ref{vector0}), which is compatible with the
recent measurements.

The scalar strangeness is special because of non-vanishing
scalar $q\bar q$ condensates of the QCD vacuum, which makes
it more natural that it is larger than vector, axial or
other strangeness channels. This is especially clear in
our approach applied to the MIT bag model, where the
scalar strangeness comes mostly from the difference of
the scalar $q\bar q$ condensates in the true QCD vacuum
and their absence in the perturbative vacuum inside the
cavity \cite{DoN86}, while only the relatively small remainder
comes from the response of the valence ground state to
the strangeness--sensitive probe.
However, such a response is all that exists in the case
of the pseudoscalar, vector, axial, {\it etc.}, nucleon
strangeness, since there are no
pseudoscalar, vector, axial, {\it etc.}, QCD-vacuum
condensates either inside or outside the cavity.
%Since there are no
%pseudoscalar, vector, axial, {\it etc.}, QCD-vacuum
%condensates either inside or outside the cavity, it
%follows that the pseudoscalar, vector, axial, {\it etc.},
%nucleon strangeness comes only from  the response of the
%valence ground state to the strangeness--sensitive probe.
Since such responses tend to be much smaller than the
nonperturbative vacuum contributions, significant
differences in magnitude between the scalar and other
kinds of strangeness are very natural in our approach.
In fact, in the present case of the MIT bag model, we find
the vanishing first--order contribution to the
vector strangeness. The vanishing first--order
contributions are found also for the pseudoscalar
and axial strangeness of the nucleon.

This confirms the conjecture of ref. \cite{Zh97}
for the case of the scalar stran\-ge\-ness. 
Our results are also consistent with the most recent
measurements of the strange
vector form factors at low momentum transfer, $Q^2 \lsim 1$ GeV.
The experimental stran\-ge magnetic form factor of the nucleon
at $Q^2 =0.1$ (GeV/c)$^2$,
$G^{s}_{M}=0.23\pm 0.37\pm 0.15 \pm 0.19\,\mu_{N}$ ,
obtained at MIT/Bates \cite{Mu97} is consistent with the
absence of strange quarks, but the error bars are large.
However, the results and conclusions of our approach, that
channels other than the scalar one should not be appreciably
affected by strange quarks, seems to get support especially from
the most recent and very precise TJNAF (JLab) measurement
\cite{An99}
yielding the small strange vector form factors at $Q^2 =0.48$
(GeV/c)$^2$,
$G^{s}_{E}+0.39 G^{s}_{M} = 0.023 \pm 0.034
 \pm 0.022 \pm 0.026\,\mu_{N}$ 

This also makes understandable
%provides an explanation/clarification
why the results on the ``non-scalar" strange quantities
such as the strangeness nucleon magnetic form factor
\cite{Ja89,PaSW91,MuB94,HoPM97,HoP93,Le96,Le95,HaMD96,Ch96}
or the strangeness electric mean-square radius
\cite{Ja89,PaSW91,MuB94,FoNJC94,HoPM97,HoP93,GeI97,
Le96,Le95,HaMD96,Ch96}
vary so much, even by the sign, from one model to
another: the ``non-scalar" strange quantities should
all be rather small, and artifacts of various
models very easily put in on either side of the zero.

\section*{Appendix: MIT bag--model wave functions}
\renewcommand{\theequation}{\mbox{A\arabic{equation}}}
\setcounter{equation}{0}

Quantum fields for quarks of flavour
$q=u, d$ or $s$ in the MIT bag model are
\begin{equation}
q(x)=\sum_K \left( \calQ_K\, q_{K}
( { \mbox{\boldmath$r$}})e^{-i\omega_K t} +
  \calQ^{c^\dagger}_K\, q_{K}^{c}
( { \mbox{\boldmath$r$}})e^{ i\omega_K t}
        \right) \;,
\end{equation}
\begin{equation}
\bar{q}(x)=\sum_K \left( \calQ^{\dagger}_K\, \bar{q}_K
    ( { \mbox{\boldmath$r$}}) e^{i\omega_K t} +
   \calQ_K^{c}\, \bar{q}^{c}_K
( { \mbox{\boldmath$r$}})e^{-i\omega_K t} \right) \;,
\end{equation}

\vspace*{1em}
\noindent where $\calQ, \calQ^{\dagger}, \calQ^{c}$ and
$\calQ^{c^\dagger}$ are
annihilation and creation operators for quarks and antiquarks, respectively.
Quark and antiquark wave functions, specified by the quantum numbers
$K=\{n,j,j_3,l\}$, are \cite{CaOW86}
\vspace*{0.5cm}
\begin{equation}
q_{n j j_3 l}( \mbox{\boldmath$r$})=N_{jl}(x_{njl})\left(
\begin{array}{c}
 i W_{+}(njl)\; j_l\left(x_{njl}\frac{r}{R}\right)\;\phi_{jj_3l}
(\hat{ \mbox{\boldmath$r$}})\\
 \\ (\bar{l}-l) W_{-}(njl)\;j_{\bar{l}}\left(x_{njl}\frac{r}{R}\right)
 \;\phi_{jj_3\bar{l}}(\hat{ \mbox{\boldmath$r$}}) \end{array}
 \right) \;,
\end{equation}
\vspace*{1cm}
\begin{equation}
q^{c}_{n j j_3 l}( \mbox{\boldmath$r$})=N_{jl}(x_{njl})\left(
\begin{array}{c}
 i W_{-}(njl)\; j_{\bar{l}}\left(x_{njl}\frac{r}{R}\right)
 \;\phi_{jj_3\bar{l}} (\hat{ \mbox{\boldmath$r$}})\\
 \\ (\bar{l}-l) W_{+}(njl)\;j_{l}\left(x_{njl}\frac{r}{R}\right)
 \;\phi_{jj_3l}(\hat{ \mbox{\boldmath$r$}}) \end{array}
 \right) \;.
\end{equation}

\noindent
 Here
\begin{equation}
  \bar{l}=j\mp\frac{1}{2} \quad \mbox{when} \quad l=j\pm\frac{1}{2} \;,
\end{equation}
and
\begin{equation}
  W_{\pm}(njl)=\sqrt{\frac{\omega_{njl}\pm m_q}{\omega_{njl}}} \;.
\end{equation}

\noindent The normalization constant is
\begin{equation}
N^{-2}_{jl}(x_{njl})=\frac{R^3 j_{l}^2(x_{njl})}
{\omega_{njl}(\omega_{njl}-m_q)}
\left\{2\omega_{njl} \left[\omega_{njl}-(\bar{l}-l)
\frac{j+\frac{1}{2}}{R}\right]
+\frac{m_q}{R}\right\} \;,
\end{equation}

\noindent and the angular parts of the wave functions are
\begin{equation}
\phi_{jj_3l}(\hat{ \mbox{\boldmath$r$}})=\sum_{l_3s_3}\,
\langle jj_3 | ll_3 , \frac{1}{2}s_3\rangle\,
Y_{l}^{l_3}(\hat{ \mbox{\boldmath$r$}})\, \chi_{s_3} \;,
\end{equation}
\begin{equation}
\phi_{jj_3\bar{l}}(\hat{ \mbox{\boldmath$r$}})=- \mbox{\boldmath$\sigma$}
\cdot \hat{\mbox{\boldmath$r$}} \phi_{jj_3l}(\hat{ \mbox{\boldmath$r$}})
\;.
\end{equation}
\noindent Here, $Y_{l}^{l_3}$ are spherical harmonics, $\chi_{s_3}$ are
Pauli spinors, $\langle jj_3 | ll_3 , \frac{1}{2}s_3\rangle$ are
Clebsch-Gordan coefficients, $\mbox{\boldmath$\sigma$}$ are Pauli matrices
and $j_l (\rho)$ are spherical Bessel functions.
$R$ is the bag radius, and $m_q$ is the quark mass. The energy eigenvalues
\begin{equation}
\omega_{njl}=\sqrt{\frac{x_{njl}^2}{R^2}+m^{2}_q} \;,
\end{equation}
are determined by the roots $x_{njl}$ of the equation
\begin{equation}
  j_l(x)=(\bar{l}-l) \sqrt{\frac{\omega_{njl} - m_q}{\omega_{njl} +
m_q}}\,j_{\bar
{l}}(x)
  = (\bar{l}-l)\frac{W_{-}}{W_{+}}\,j_{\bar{l}}(x) \;.
\end{equation}

\noindent Instead of $\{j,l\}$ we can use the quantum number $\{\kappa\}$
such that
\begin{equation}
j=|\kappa |-\frac{1}{2} \;,
\end{equation}
and
\begin{equation}
l=|\kappa |+\frac{\textrm{sign}(\kappa)-1}{2} \;,
\end{equation}
\begin{equation}
\bar{l}=|\kappa |-\frac{\textrm{sign}(\kappa)+1}{2} \;.
\end{equation}
In this case, the wave functions are specified by the quantum
numbers $K=\{n,\kappa, j_3\}$ and are of the form

\begin{equation}
q_{n \kappa j_3}( \mbox{\boldmath$r$})=N_{\kappa}(x_{n\kappa})\left(
\begin{array}{c}
 i W_{+}(n\kappa)\;
j_l\left(x_{n\kappa}\frac{r}{R}\right)\;\phi_{\kappa}^{j_3}
(\hat{ \mbox{\boldmath$r$}})\\
 \\ -\textrm{sign}(\kappa) W_{-}(n\kappa)\;j_{\bar{l}}\left(x_{n\kappa}
\frac{r}{R}\right)
 \;\phi_{-\kappa}^{j_3}(\hat{ \mbox{\boldmath$r$}}) \end{array}
 \right) \;,
\end{equation}
\vspace*{1cm}
\begin{equation}
q^{c}_{n \kappa j_3}( \mbox{\boldmath$r$})=N_{\kappa}(x_{n\kappa})\left(
\begin{array}{c}
 i W_{-}(n\kappa)\; j_{\bar{l}}\left(x_{n\kappa}\frac{r}{R}\right)
\;\phi_{-\kappa}^{j_3}
(\hat{ \mbox{\boldmath$r$}})\\
 \\ -\textrm{sign}(\kappa) W_{+}(n\kappa)\;j_{l}\left(x_{n\kappa}
\frac{r}{R}\right)
 \;\phi_{\kappa}^{j_3}(\hat{ \mbox{\boldmath$r$}}) \end{array}
 \right) \;,
\end{equation}
where the normalization constant is
\begin{equation}
N^{-2}_{\kappa}(x_{n\kappa})=\frac{R^3
j_{l}^2(x_{n\kappa})}{\omega_{n\kappa}(\omega_{n\kappa}-m_q)}
\left\{2\omega_{n\kappa} \left[\omega_{n\kappa}+\frac{\kappa}{R}\right]
+\frac{m_q}{R}\right\} \;,
\end{equation}
the angular parts of the wave functions are
\begin{eqnarray}
\phi_{\kappa}^{j_3}(\hat{ \mbox{\boldmath$r$}}) &=&
-\textrm{sign}(\kappa)\sqrt{\frac{|\kappa|+\textrm{sign}(\kappa)
(\frac{1}{2}-j_3)}{2|\kappa|+\textrm{sign}(\kappa)}}\,Y^{j_3-\frac{1}{2}}_l
(\hat{ \mbox{\boldmath$r$}})
\chi^{\frac{1}{2}} \nonumber \\
 &&+ \sqrt{\frac{|\kappa|+\textrm{sign}(\kappa)
(\frac{1}{2}+j_3)}{2|\kappa|+\textrm{sign}(\kappa)}}\,Y^{j_3+\frac{1}{2}}_l
(\hat{ \mbox{\boldmath$r$}})
\chi^{-\frac{1}{2}} \;,
\end{eqnarray}

\noindent and $\omega_{n\kappa}$'s are given by the equation
\begin{equation}
  j_l(x)+ \textrm{sign}(\kappa)\frac{W_{-}}{W_{+}}\,j_{\bar{l}}(x)=0 \;.
\end{equation}
Our conventions follow those of \cite{CaOW86}.
%%%% end of Appendix
\renewcommand{\theequation}{\arabic{equation}}

\subsection*{Acknowledgment}

{\small D. K. and I. P. thank I. Zahed for getting them
started in this problem, and for many illuminating discussions.}

%When BiBTeX is used:
%\bibliographystyle{h-physrev3}
%\bibliography{kkmp98refs}

\end{document}